\documentclass{aa}

\usepackage{graphicx}
\usepackage{color}
\usepackage{amssymb}
\usepackage[version=4]{mhchem}
\usepackage[para]{threeparttable}

\usepackage{txfonts}

\begin{document} 

\title{High D/H ratios in water and alkanes in comet 67P/Churyumov-Gerasimenko measured with the Rosetta/ROSINA DFMS}


\author{D. R. M\"uller\inst{1} \and K. Altwegg\inst{1} \and J. J. Berthelier\inst{2} \and M. Combi\inst{3} \and J. De Keyser\inst{4} \and S. A. Fuselier\inst{5,6} \and N. H\"anni\inst{1} \and B. Pestoni\inst{1} \and M. Rubin\inst{1} \and I. R. H. G. Schroeder I\inst{1} \and S. F. Wampfler\inst{7}
      }

\institute{	Physikalisches Institut, University of Bern, Sidlerstrasse 5, 3012 Bern, Switzerland\\
          \email{daniel.mueller@unibe.ch}
     \and
            Laboratoire Atmosph\`eres, Milieux, Observations Spatiales (LATMOS), 4 Avenue de Neptune, 94100 Saint-Maur, France
     \and
            Department of Climate and Space Sciences and Engineering, University of Michigan, 2455 Hayward, Ann Arbor, MI 48109, USA
     \and
            Royal Belgian Institute for Space Aeronomy, BIRA-IASB, Ringlaan 3, 1180 Brussels, Belgium
     \and
            Space Science Directorate, Southwest Research Institute, 6220 Culebra Rd., San Antonio, TX 78228, USA
     \and
            Department of Physics and Astronomy, The University of Texas at San Antonio, San Antonio, TX 78249, USA
     \and
         	Center for Space and Habitability, University of Bern, Gesellschaftsstrasse 6, 3012 Bern, Switzerland\\
         }

\date{Received 16 December 2021 / Accepted 6 February 2022}

\abstract
{Isotopic abundances in comets are key to understanding and reconstructing the history and origin of material in the Solar System. Data for deuterium-to-hydrogen (D/H) ratios in water are available for several comets. However, no long-term studies of the D/H ratio in water of a comet during its passage around the Sun have been reported. Linear alkanes are important organic molecules, which have been found on several Solar System bodies, including comets. To date, their deuteration is still poorly understood, as only upper limits of isotopic ratios for D/H and \textsuperscript{13}C/\textsuperscript{12}C in linear alkanes are available.}
{The aim of this work is a detailed analysis of the D/H ratio in water as a function of cometary activity and spacecraft location above the nucleus. In addition, a first determination of the D/H and \textsuperscript{13}C/\textsuperscript{12}C ratios in the first four linear alkanes, namely, methane (CH\textsubscript{4}), ethane (C\textsubscript{2}H\textsubscript{6}), propane (C\textsubscript{3}H\textsubscript{8}), and butane (C\textsubscript{4}H\textsubscript{10}) in the coma of 67P/Churyumov-Gerasimenko is provided.}
{We analysed in situ measurements from the Rosetta/ROSINA Double Focusing Mass Spectrometer (DFMS).}
{The D/H ratio from HDO/H\textsubscript{2}O and the \textsuperscript{16}O/\textsuperscript{17}O ratio from H\textsubscript{2}\textsuperscript{16}O/H\textsubscript{2}\textsuperscript{17}O did not change during 67P's passage around the Sun between 2014 and 2016. All D/H ratio measurements were compatible, within 1$\sigma$, with the mean value of $5.01\times10^{-4}$ and its relative variation of 2.0\%. This suggests that the D/H ratio in 67P's coma is independent of heliocentric distance, level of cometary activity, as well as spacecraft location with respect to the nucleus. Additionally, the \textsuperscript{16}O/\textsuperscript{17}O ratio could be determined with a higher accuracy than previously possible, yielding a value of 2347 with a relative variation of 2.3\%. 
For the alkanes, the D/H ratio is between 4.1 and 4.8 times higher than in H\textsubscript{2}O, while the \textsuperscript{13}C/\textsuperscript{12}C ratio is compatible, within uncertainties, with data for other Solar System objects. The relatively high D/H ratio in alkanes is in line with other cometary organic molecules and suggests that these organics may be inherited from the presolar molecular cloud from which the Solar System formed.}
{}

\keywords{Comets: general –- Comets: individual: 67P/Churyumov-Gerasimenko -- Instrumentation: detectors -- Methods: data analysis -- Astrochemistry}
\titlerunning{High D/H ratios in water and alkanes in comet 67P}
\maketitle
%
\section{Introduction}\label{sec:introduction}
Comets are considered to be reservoirs of material preserved from the early Solar System. By making this material available to in-situ exploration, cometary science contributes important information on the history of the Solar System \citep{Drozdovskaya2019, Mumma2011}.
Investigating the isotopic abundances of different elements in various molecules in comets is essential, as the isotopic ratios are sensitive to the environmental conditions at the time of the molecules' formation and provide crucial information for the understanding of the origins of cometary material \citep{Biver2019, Bockelee2015, Haessig2017}.

The best-studied comet to date is comet 67P/Churyumov-Gerasimenko (hereafter 67P), a Jupiter-family comet (JFC). 67P was followed by the Rosetta spacecraft during its orbit around the Sun. In August 2014, Rosetta rendezvoused with 67P at a heliocentric distance of around 3.6 au. It then accompanied the comet through its perihelion at 1.24 au from the Sun and followed the orbit of 67P back out to a distance of almost 4 au, whereupon the spacecraft intentionally soft-landed on the comet’s surface at the end of September 2016. The Rosetta spacecraft was launched and operated by the European Space Agency (ESA). The Rosetta mission uncovered a lot of new knowledge about 67P, such as its gas and dust composition (e.g. \citealp{Herny2021, Longobardo2020, Pestoni2021}), nucleus surface (e.g. \citealp{Feller2019}) and temporal evolution (e.g. \citealp{Combi2020, Laeuter2020, Rubin2019}). With its lander, Philae, it was even able to acquire gas and volatiles in dust composition data directly on or near the comet's surface by the COSAC \citep{Goesmann2015} and Ptolemy \citep{Wright2015} instruments. No prior cometary observation has ever been performed for as long a duration and with as high a measurement sensitivity as the Rosetta mission.

The Rosetta spacecraft carried several onboard instrument packages. One of them was the Rosetta Orbiter Spectrometer for Ion and Neutral Analysis (ROSINA). ROSINA was comprised of two mass spectrometers, the Double Focusing Mass Spectrometer (DFMS) and a Reflectron-type Time-Of-Flight mass spectrometer (RTOF), and a COmet Pressure Sensor (COPS). DFMS was used for measurements of the molecular and isotopic composition of cometary volatiles \citep{Balsiger2007}. \citet{Haessig2017} showed that the instrument had a sensitivity, dynamic range and mass resolution high enough to detect even trace amounts of rare isotopologues alongside their more abundant counterparts. It was used by many authors to investigate the isotopic ratios of sulfur \citep{Calmonte2017, Haessig2017}, carbon \citep{Haessig2017, Altwegg2020}, the halogens bromine and chlorine \citep{Dhooghe2017}, and oxygen \citep{Altwegg2020, Haessig2017, Schroeder2019b} in 67P. \cite{Altwegg2015, Altwegg2017} used it to measure the D/H ratio in water in 67P’s coma, using data from the beginning and near the end of the Rosetta mission. Both measurements were consistent within uncertainties. From HDO/H\textsubscript{2}O, a D/H ratio of $(5.3 \pm 0.7)\times10^{-4}$ was deduced. This is more than three times the terrestrial Vienna Standard Mean Ocean Water (VSMOW) value of $1.5576\times10^{-4}$, and one of the highest ever measured in a JFC.

Both measurements by \cite{Altwegg2015, Altwegg2017} were performed at times when 67P was relatively far from the Sun. The first had relied on data from well before perihelion, in August/September 2014 at a heliocentric distance of 3.4 au, while the second evaluated data from December 2015 at 2 au and the outbound equinox in March 2016 at 2.6 au. Due to the large heliocentric distances of 67P during these measurements, the question arises as to whether the HDO/H\textsubscript{2}O ratio would differ at smaller heliocentric distances, when a large increase in sublimation from the surface of the cometary nucleus occurred and fresh layers of the comet’s surface were likely exposed. Additionally, different cometary hemispheres were active at different times. At greater heliocentric distances, most of the water outgassed came from the comet’s northern latitudes. Conversely, closer to perihelion, the contributions of the southern latitudes were more significant \citep{Keller2015}. \citet{Schroeder2019a} investigated the difference between the comet's two lobes and concluded that no significant difference in the D/H ratio could be observed.

\begin{figure}
    \centering
    \includegraphics[width=\hsize]{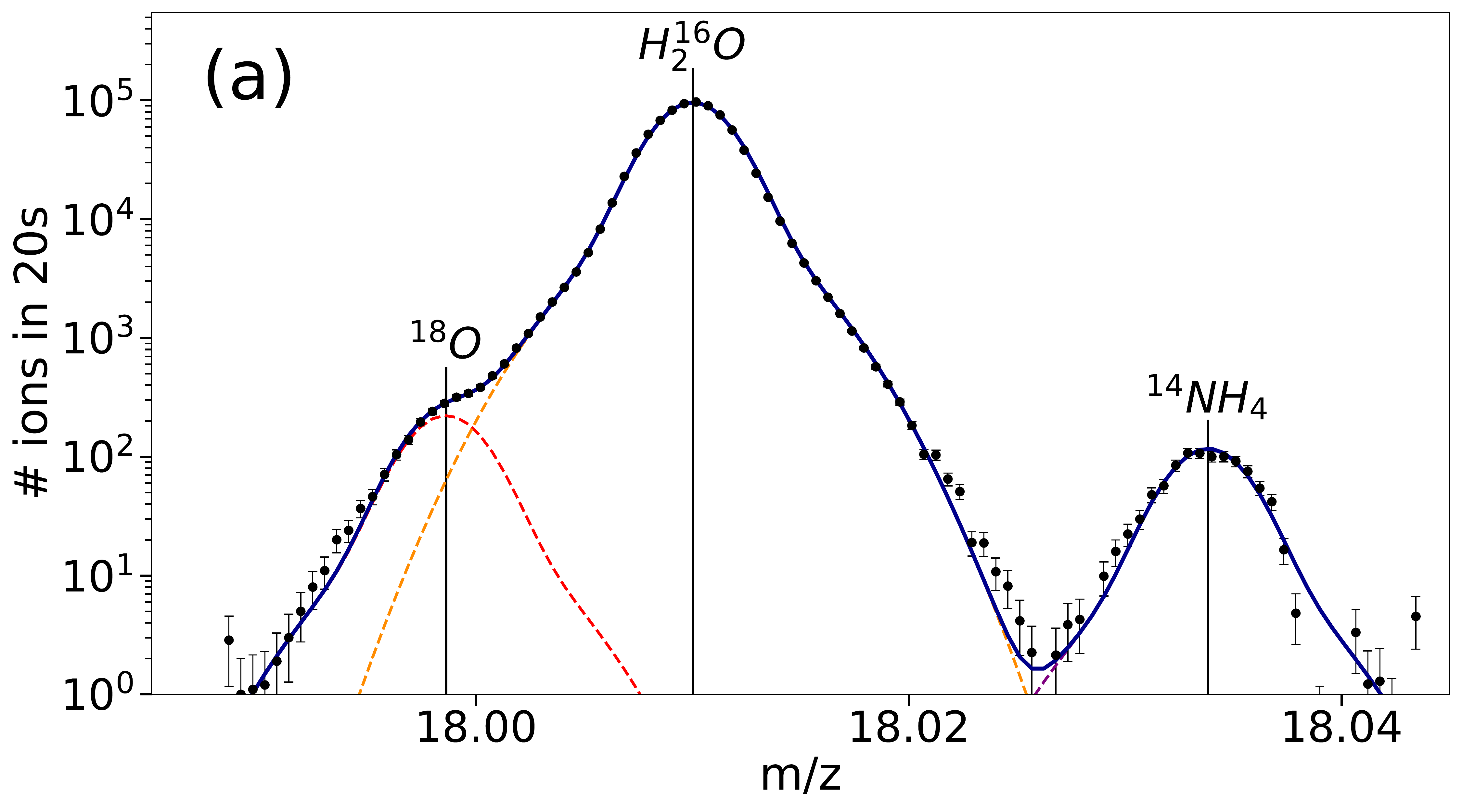}\hfill
    \includegraphics[width=\hsize]{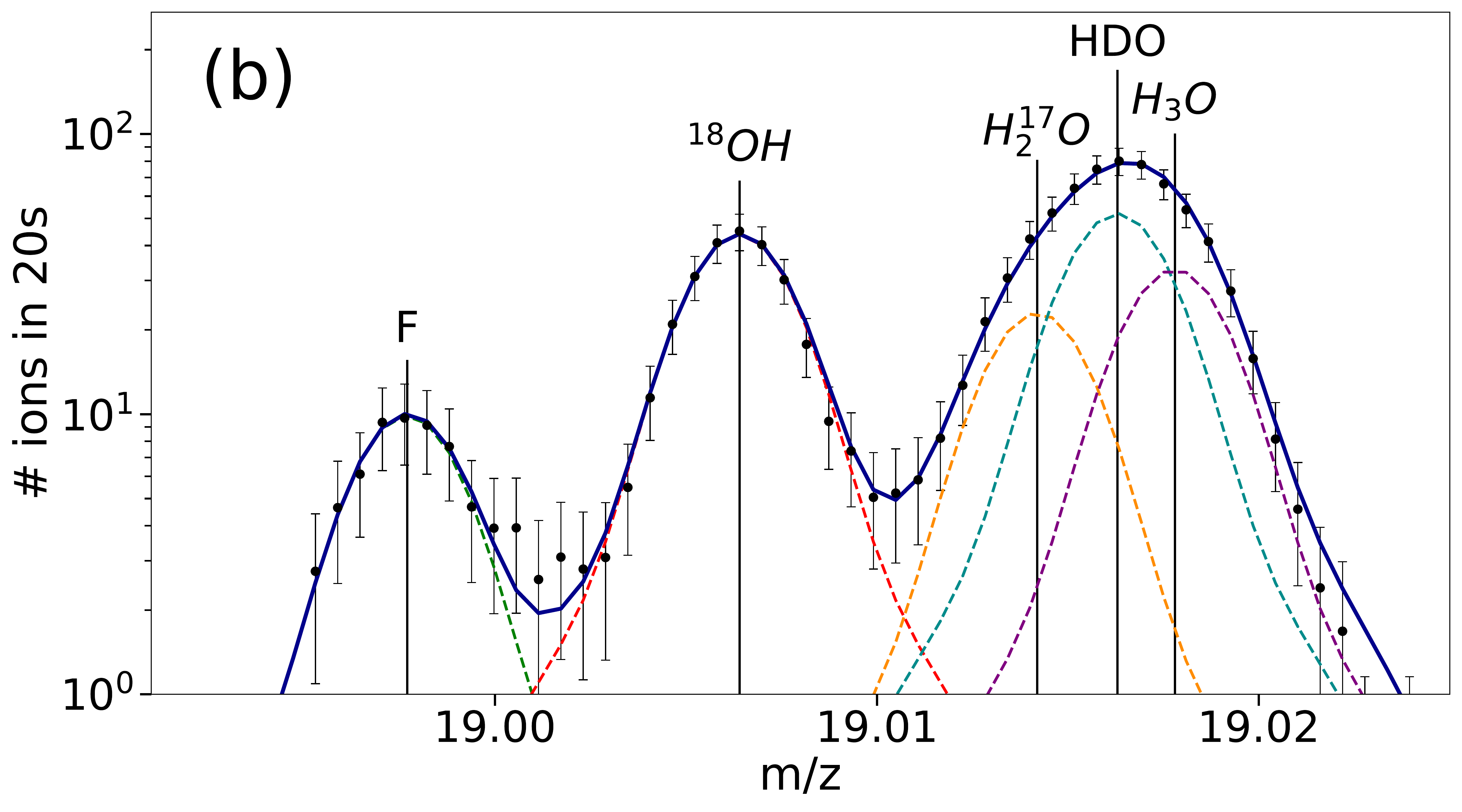}\hfill
    \caption{Sample mass spectra for \emph{m/z} 18 and 19 showing the signatures of the isotopologues of water. \textit{Panel a}: \emph{m/z} 18 from 2015-05-07 17:38 (UTC). \textit{Panel b}: \emph{m/z} 19 from 2015-05-26 01:35 (UTC). Measured data are represented by black dots including their statistical uncertainties. Individual mass fits and the total sum of the fits are shown with coloured lines.}
    \label{fig:sampleSpectraWater}
\end{figure}

A comparison of different Solar System objects shows a large variation in D/H ratios, with most objects being enriched in deuterium compared to the protosolar nebula \citep{Altwegg2015}. Different potential mechanisms have been proposed to explain these large variations, for instance solar wind induced water formation and isotopic fractionation. Daly et. al (2021) states, that isotopically light water reservoirs could have been produced by solar wind implantation into fine-grained silicates. The authors conclude that this may have been a particularly important process in the early Solar System and thus potentially provides a means to recreate Earth’s current water isotope ratios. On the other hand, the isotopic fractionation describes the variation in abundances of the isotopes of an element. It arises from both physical and chemical processes and is also temperature-dependent for some molecules. According to \citet{Kavelaars2011}, the main reservoir of deuterium in the protosolar nebula was molecular hydrogen with a D/H ratio of $1.5\times10^{-5}$. Ion-molecule reactions in the interstellar medium or grain surface chemistry can cause fractionation among deuterated species. In the pre-solar cloud, fractionation resulted in molecules being enriched in deuterium. Isotopic exchange reactions with H\textsubscript{2} in the gas phase of the solar nebula would then lower this enrichment. Various authors suggested that the enrichment in deuterium increases with increasing heliocentric distance \citep{Furuya2013, Kavelaars2011, Geiss1981}. Comets are assumed to be a source of primordial material from the early Solar System \citep{Wyckoff1991}. Consequently, knowledge of variations in the deuterium enrichment in comets is of high importance, as their compositions are indicative of their regions of origin and the environmental conditions during their formation \citep{Haessig2017}.

Ground-based observations of deuterated water in comet C/2014 Q2 (Lovejoy), hereafter Lovejoy, appeared to show a change in the D/H ratio in water from pre- to post-perihelion \citep{Paganini2017}. \citet{Paganini2017} measured a post-perihelion D/H ratio of $(3.02 \pm 0.87)\times10^{-4}$, which was significantly higher than the pre-perihelion value of $(1.4 \pm 0.4)\times10^{-4}$ measured by \citet{Biver2016}. Two explanations for this discrepancy were put forward by \citet{Paganini2017}: (1) The ratio of D/H in water changed after perihelion, or (2) the D/H ratio in water might have been strongly influenced by a systematic bias in the estimate as different experimental setups were used. \citet{Paganini2017} used the Near Infrared Spectrograph (NIRSPEC) at the 10-meter W. M. Keck Observatory (Keck II) for their infrared measurements. In contrast, \citet{Biver2016} used radio/sub-mm observations from the IRAM 30 m radio telescope and the Odin 1.1 m submillimeter satellite. The two different approaches and the use of two telescopes with different beam sizes in the measurements by \citet{Biver2016} could provide a possible explanation for the varying D/H results in this comet.

A recent study of the D/H ratio in comets showed that the D/H ratio correlates with the nucleus’ active area fraction \citep{Lis2019}. According to the authors' definition, comets with an active fraction larger than 0.5 are called hyperactive comets and typically exhibit D/H ratios in water consistent with the terrestrial value. The authors argue that these hyperactive comets require an additional source of water vapour within their comae, which might be explained by the presence of subliming icy grains ejected from the nucleus. There exist other definitions of hyperactivity in comets, such as in \citet{Sunshine2021}, and thus the classification of 67P to be hyperactive or not is not always clear. 
\citet{Fulle2021} hypothesizes that the correlation of the D/H ratio with the nucleus’ active area fraction might be due to a mixture of water-rich and water-poor pebbles. The author states that the two kinds of pebbles contain different D/H ratio values due to their initial formation conditions. He also suggests that the D/H average in the nuclei may differ from the values measured in cometary comae, and can therefore not be obtained by local sample-return missions. According to the author, cryogenic return missions would sample water-rich and water-poor pebbles separately, which would only be representative of their corresponding water-rich or water-poor regions, respectively. A cometary average therefore cannot be measured by local sampling.

This work is the first to assess the scenario of a changing D/H ratio in a comet with numerous data points from in-situ measurements spread over a long time period and shall answer the question of whether the D/H ratio in comets is dependent on heliocentric distance, phase angle or gas production rate. To do so, we evaluated the full mission data of ROSINA/DFMS to investigate the D/H ratio in HDO and H\textsubscript{2}O over one third of 67P's orbit. The evaluated mission phases are specified in Table~\ref{table:H2OSummary}.

In addition to water, the D/H ratios of different alkanes have been studied. Alkanes are acyclic saturated hydrocarbon molecules containing only single carbon-carbon bonds. They have been found on several Solar System bodies, including the Earth, and in the atmospheres of the giant planets and Saturn's moon Titan \citep{Clark2009, Lunine2008}. The isotopic ratios in these organic compounds are of special interest as they may provide not only an insight into the chemical and physical conditions before and during the formation of the Solar System, but can also constrain the delivery of organic matter by comets to the early Earth \citep{Doney2020, Rubin2019, Schuhmann2019}.

\section{Instrumentation \& Methodology}
\begin{figure}
    \centering
    \includegraphics[width=\hsize]{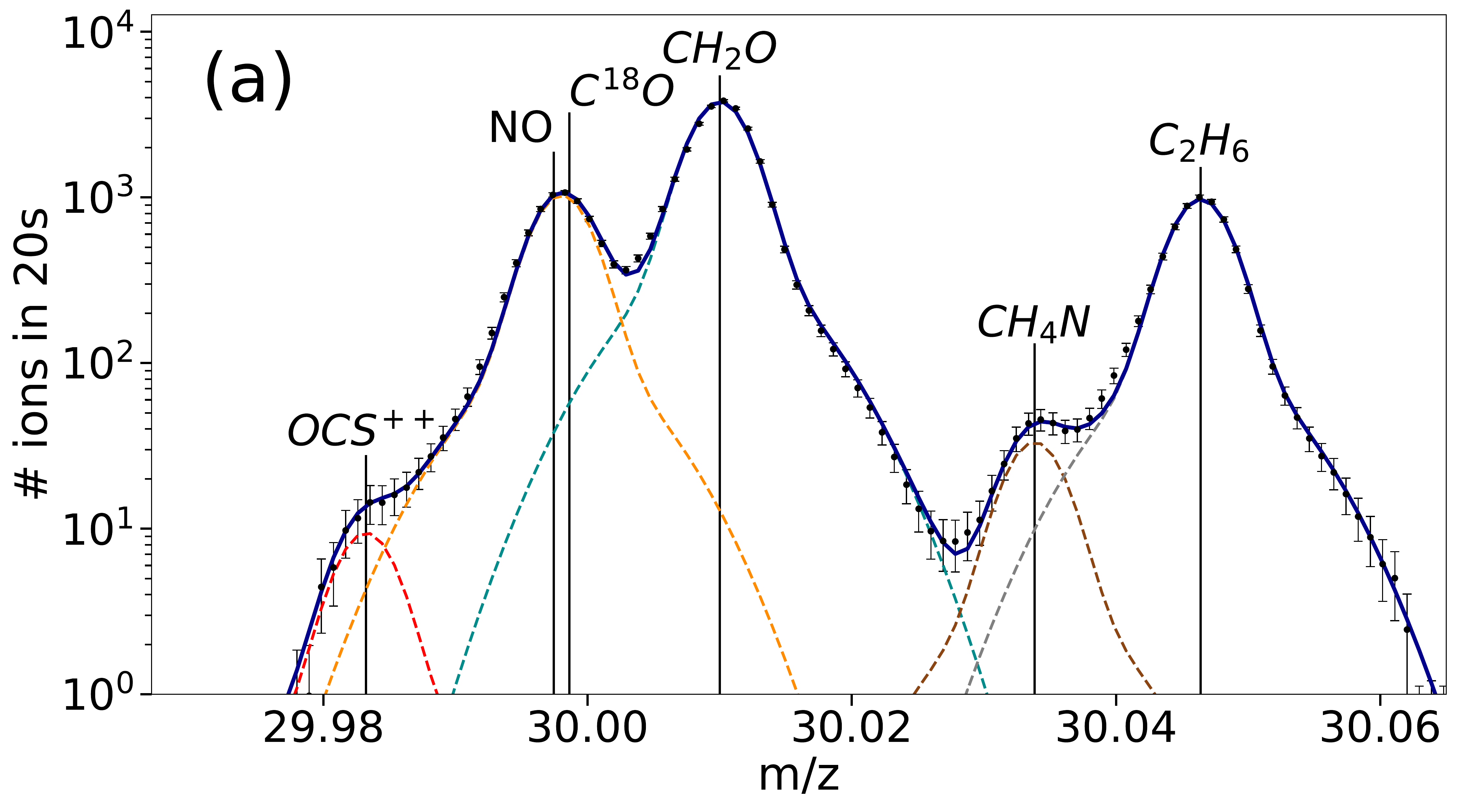}\hfill
    \includegraphics[width=\hsize]{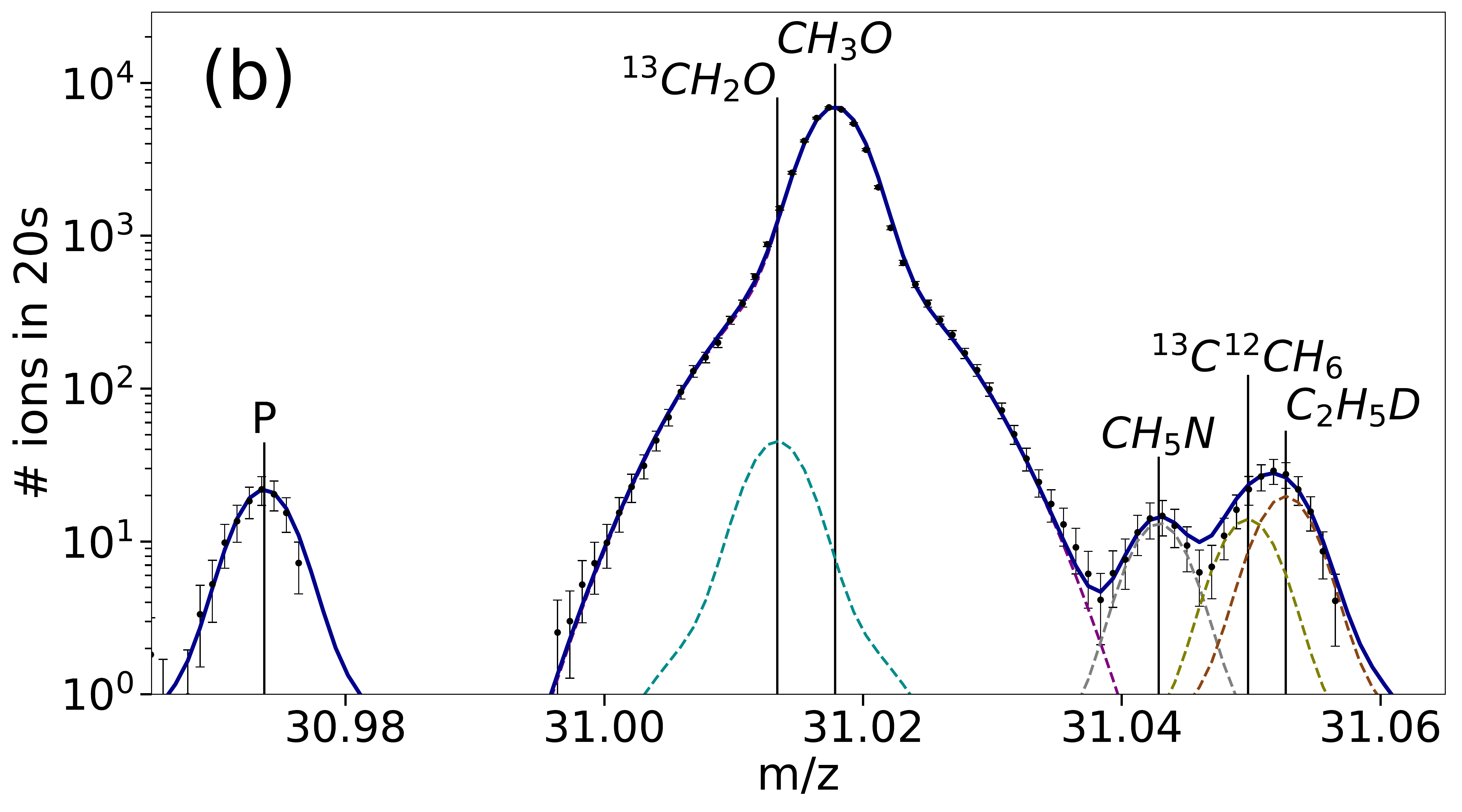}
    \caption{Sample mass spectra for \emph{m/z} 30 and 31 showing the signatures of the isotopologues of ethane. \textit{Panel a}: \emph{m/z} 30 from 2014-03-10 19:36 (UTC). The peaks of NO and C\textsuperscript{18}O could not be resolved and appear as one peak (orange line). \textit{Panel b}: \emph{m/z} 31 from 2014-03-10 19:37 (UTC). Measured data are represented by black dots including their statistical uncertainties. The individual mass fits and the total sum of the fits are shown with coloured lines.}
    \label{fig:sampleSpectraAlkanes}
\end{figure}
The ROSINA/DFMS is a Nier-Johnson type double focusing mass spectrometer with a high mass resolution of m/$\Delta$m = 3000 at the 1\%-level on the mass-to-charge ratio (\emph{m/z}) 28 \citep{Balsiger2007}. In the DFMS, incoming neutral gas is ionized by electron impact with an electron energy of 45 eV. Most ions formed are singly charged. For this reason, the charge state will not be indicated in the following, except for the subset of doubly charged ions, such as H\textsubscript{2}S\textsuperscript{++}. The newly formed ions are accelerated through a 14-µm slit, deflected by 90 degrees in a toroidal electrostatic analyser and finally undergo a 60-degree deflection in the field of a permanent magnet. With the combination of the different fields, the instrument is tuned such that only ions with a specific mass-to-charge ratio make it through the analyser section. The remaining ion beam is amplified by two micro channel plates (MCP) in a Chevron configuration. The electron packet issued from the MCP is finally collected by a position-sensitive Linear Electron Detector Array (LEDA). The LEDA consists of two rows with 512 pixels each \citep{Nevejans2002}.

The MCP potential difference can be varied to adjust its amplification. The amplification is the gain of the MCP. 16 different settings or gain steps can be chosen from default voltages. Due to detector ageing, the gains associated with each voltage settings are not constant over time. This has to be accounted for when comparing DFMS data with different gain steps. In addition, the unequal usage of the 512 pixels of the LEDA causes a position-dependent degradation of the detector over time. For this reason, a pixel gain correction needs to be implemented during data evaluation \citep{DeKeyser2019}. Gain and pixel gain correction factors were evaluated in \citet{Schroeder2019b}.

\begin{table*}
\caption{D/H in H\textsubscript{2}O during different mission phases and compared to previous evaluations.}             
\label{table:H2OSummary}      
\centering
\begin{tabular}{ p{3cm}p{3.5cm}p{3cm}p{2cm}p{2cm}  }
\hline
\noalign{\smallskip}
Mission Phase & Dates & D/H in H\textsubscript{2}O & Heliocentric distance [au] & \# of evaluated spectra\\
\noalign{\smallskip}
\hline
\noalign{\smallskip}
1st Equinox & May 2015 & (5.03 ± 0.17) · 10\textsuperscript{-4} & 1.71 - 1.52 & 44\\
Perihelion & August 2015 & (5.01 ± 0.20) · 10\textsuperscript{-4} & 1.24  & 37\\
Peak gas production & 2015-08-30 & (4.98 ± 0.25) · 10\textsuperscript{-4} & 1.26 & 22\\
2nd Equinox & March 2016 & (5.02 ± 0.17) · 10\textsuperscript{-4} & 2.45 - 2.65 & 47\\
\noalign{\smallskip}
\hline
\noalign{\smallskip}
Relative mean ratio && (5.01 ± 0.10) · 10\textsuperscript{-4} &  & 150\\
Absolute mean ratio  && (5.01 ± 0.41) · 10\textsuperscript{-4} &  & 150\\
\noalign{\smallskip}
\hline                
\noalign{\smallskip}
\hline
\noalign{\smallskip}
Pre-1st Equinox \textsuperscript{a} & Aug. / Sep. 2014 & (5.3 ± 0.7) · 10\textsuperscript{-4} & $\approx$3.4 & 26\\
Pre-2nd Equinox \textsuperscript{b} & Dec. 2015 / Mar 2016 & (5.25 ± 0.7) · 10\textsuperscript{-4} & 2.0 \& 2.6 & 18\\
\noalign{\smallskip}
\hline
\end{tabular}
\tablebib{
(a)~\citet{Altwegg2015}; (b)~\citet{Altwegg2017}.
}
\end{table*}

A single spectrum comprises a range of \emph{m/z} around a specified integer \emph{m/z}. For \emph{m/z} 28, this is ±0.25. DFMS spectra are fitted on individual mass lines using the sum of two Gaussian peaks (double Gaussian distribution). The second Gaussian depends on the first one as its signal amplitude is approximately 10\% of the first Gaussian and its width is about 3 times broader than the narrow first Gaussian. All peaks on the same spectrum are characterized by the same width and height ratios of the two Gaussian distributions. The inter-dependence of the two Gaussians is known from thorough calibration measurements by \citet{LeRoy2015} and \citet{Haessig2013, Haessig2015}, wherein the combined influence of the molecular ionization cross-sections, the mass-dependent instrument transfer function, isotope-dependent fractionation patterns due to the electron impact ionization, and detector yields have been investigated. Their effects are included in the systematic error calculations.

Finally, a mass scale may be applied to the spectrum such that each pixel corresponds to a certain mass. The mass scale is applied as described in detail in \citet{Calmonte2016}.

Formally, for each pixel $p_i$ corresponding to a LEDA pixel in the DFMS mass spectrum, the counted number of particles, counts(p\textsubscript{i}), can be described as:
\begin{equation}
      \mathrm{counts(p_i)} = a_1 e^{-(\frac{p_i-p_0}{c_1})^2} + a_2 e^{-(\frac{p_i-p_0}{c_2})^2} \,,
   \end{equation}
with $a_1$ and $a_2$ being the amplitudes of the first and the second Gaussian, respectively, $p_0$ the pixel zero corresponding to the integer mass (center pixel), and $c_1$ and $c_2$ the widths of the two Gaussians.
The total number of particles impinging on the detector is represented by the peak area. It is given by the integral of the fitted double-Gaussian distribution:
\begin{equation}
      \mathrm{\#\ \,of\,particles} = \int_{-\infty}^{+\infty} \mathrm{counts(p)\,dp} = \sqrt{\pi} (a_1 c_1+a_2 c_2). \,
   \end{equation}

Figures~\ref{fig:sampleSpectraWater} and \ref{fig:sampleSpectraAlkanes} show examples of fitted mass spectra after application of the mass scale. The error bars show the statistical uncertainty on the count number.

Data from different periods during the Rosetta mission have been investigated. The HDO/H\textsubscript{2}O ratio in 67P has been examined at the first equinox (May 2015), at perihelion (August 2015), at the time of the peak gas production (end of August / early September 2015), and at the second equinox (March 2016) of 67P. These characteristic time periods have been chosen in order to determine a potential heliocentric distance dependence on the HDO/H\textsubscript{2}O ratio.

In addition, the D/H ratios of the simplest four linear alkanes - methane (CH\textsubscript{4}), ethane (C\textsubscript{2}H\textsubscript{6}), propane (C\textsubscript{3}H\textsubscript{8}) and butane (C\textsubscript{4}H\textsubscript{10}) - have been studied at times when the alkane signals were clearly visible in the respective spectra. Butane has two structural isomers, n-butane and iso-butane, that have the same molecular formula, but with the atoms in a different order. They cannot be distinguished from each other with the DFMS and thus no distinction is made in the following.
Methane and ethane have previously been detected in several comets (C/1996 B2 (Hyakutake): \citealp{Mumma1996}; 153P/Ikeya-Zhang: \citealp{Kawakita2003}; C/2007 N3 (Lulin): \citealp{Gibb2012}) and upper limits for their D/H ratios have been reported \citep{Bonev2009, Doney2020}. Propane and butane were first detected in 67P by \citet{Schuhmann2019}. These authors have also published the relative abundances of the simplest four linear alkanes compared to methane and water in 67P's coma for two different time periods. The abundances relative to water are shown in Table~\ref{table:Abundances}. The abundance of the simplest four linear alkanes strongly increased from pre- to post-perihelion. No D/H ratios for any of the alkanes considered have been reported to date.

\begin{table}
\caption{Relative abundance of alkanes in 67P. Data from \citet{Schuhmann2019}.}
\label{table:Abundances}      
\centering          
\begin{tabular}{l l l}     
\hline
\noalign{\smallskip}
Species &  \multicolumn{2}{l}{Abundance relative to water [H\textsubscript{2}O]}\\
\noalign{\smallskip}
\hline
\noalign{\smallskip}
{}   & May 2015   & May 2016\\
\noalign{\smallskip}
\hline
\noalign{\smallskip}
Methane   & (3.43 $\pm$ 0.68) · $10^{-3}$ & (6.48 $\pm$ 1.30) · $10^{-2}$\\
Ethane   & (2.92 $\pm$ 0.58) · $10^{-3}$ & (5.13 $\pm$ 1.03) · $10^{-1}$\\
Propane   & (1.80 $\pm$ 0.36) · $10^{-4}$ & (2.75 $\pm$ 0.55) · $10^{-2}$\\
Butane   & not detected & (5.28 $\pm$ 1.06) · $10^{-3}$\\
\noalign{\smallskip}
\hline                    
\end{tabular}
\end{table}

Two sources of uncertainty are relevant for DFMS data ana\-ly\-sis: statistical uncertainties in the count rates and systematic uncertainties due to instrumental effects. The statistical uncertainties of the detector counts are proportional to $\sqrt{N}$ for N counts. Additionally, a fitting error has been included in the case of overlapping peaks. This fitting error accounts for a possible ambiguity when peaks cannot be clearly separated and depends on the relative peak intensities and the mass difference between the peaks. 
Instrumental effects, arising from pixel-dependent degradation (pixel gain correction) and changes in the detector gain over time, are systematic uncertainties. The uncertainty of the pixel gain is 5\% and the uncertainty of the overall gain is 6\%. These values were previously derived and used by \citet{Schroeder2019b}. Statistical and fitting uncertainties are considered for each individual measurement point. Uncertainties in the detector and pixel gain, which are of a systematic nature, are only considered for the absolute mean ratios.
In the case of HDO/H\textsubscript{2}O, the overall gain has a large impact on the evaluation as \emph{m/z} 18 has always been measured on a smaller gain step than \emph{m/z} 19. For the alkanes, the gain uncertainty has to be included for methane and propane. The isotopologues of ethane and butane on the other hand were measured on the same gain step as their main isotopologues and gain corrections are therefore unnecessary. The pixel gain uncertainty, however, applies to all uncertainty calculations.

\section{Results}
\begin{figure*}
\centering
\includegraphics[width=\hsize]{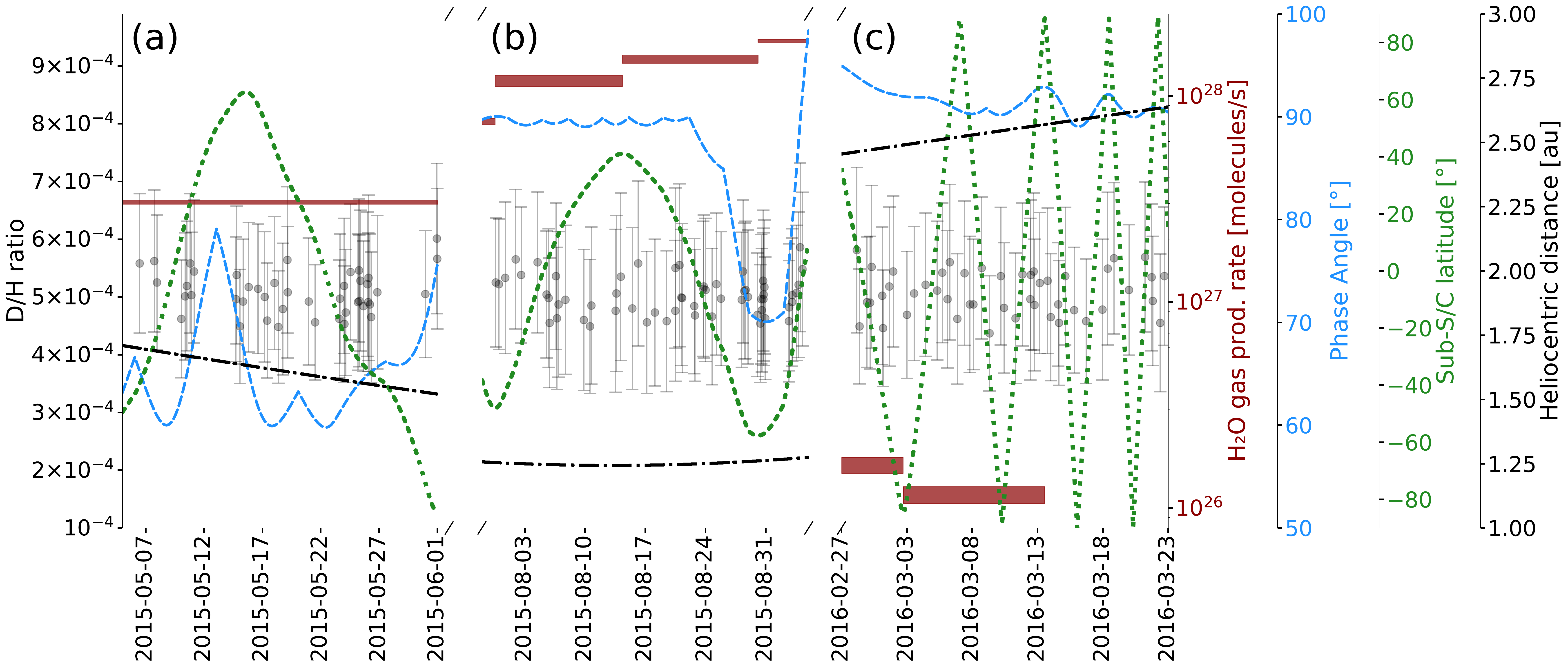}
\caption{D/H in H\textsubscript{2}O during different mission phases compared to H\textsubscript{2}O gas production (\citealp{Laeuter2020}, red), phase angle (blue), sub-S/C latitude (green) and heliocentric distance (black). \textit{Panel a}: 1\textsuperscript{st} Equinox; \textit{Panel b}: Perihelion and peak gas production phase; \textit{Panel c}: 2\textsuperscript{nd} Equinox.
The individual measurement uncertainties represent statistical errors from the count rates and errors from the fit.}
\label{fig:H2OTime}
\end{figure*}

An extensive analysis of spectra with \emph{m/z} 18 and \emph{m/z} 19 in the coma of 67P showed constant D/H and \textsuperscript{16}O/\textsuperscript{17}O ratios in water during the comet’s course around the Sun in 2015 and 2016. Furthermore, the D/H and \textsuperscript{13}C/\textsuperscript{12}C ratios in the simplest four linear alkanes could be resolved. This section summarizes the results for each of the aforementioned ratios and explains how the results have been obtained.

\subsection{HDO/H\textsubscript{2}O} \label{HDO/H2O}
A total of 150 spectra around \emph{m/z} 18 and \emph{m/z} 19 have been investigated. These spectra contain the signatures of H\textsubscript{2}\textsuperscript{16}O, H\textsubscript{2}\textsuperscript{17}O and HDO. Sample spectra for \emph{m/z} 18 and \emph{m/z} 19 are shown in Fig.~\ref{fig:sampleSpectraWater}. Using the values from H\textsubscript{2}\textsuperscript{16}O and HDO, measured  back-to-back  within  one  minute, allows us to derive the D/H ratio from HDO/H\textsubscript{2}O as:
\begin{equation}
      \mathrm{D/H} = \mathrm{\frac{1}{2} \frac{n_{\ce{HDO}}}{n_{\ce{H_{2}}\ce{^{16}O}}}}. \,
   \end{equation}

The goal of this work was to investigate the D/H ratio in water over the whole mission. Therefore, Rosetta data from the first equinox, perihelion, the time of the peak gas production, and the second equinox have been evaluated as specified in Table~\ref{table:H2OSummary}. These data sets span a wide range of heliocentric distances, observational phase angles, water production rates and sub-spacecraft (sub-S/C) latitudes.

The relative mean D/H ratios for the specified mission phases, considering only statistical and fit uncertainties, are shown in Table~\ref{table:H2OSummary}. The mean values are weighted means with the weight for each individual point being inversely proportional to its statistical uncertainty. This improves the results by giving more weight to more precise measurement points. The relative overall mean value was found to be $(5.01 \pm 0.10)\times10^{-4}$. 
All periods are consistent with this mean value within the 1$\sigma$ uncertainty of 2.0\%. There is no observable trend between the periods in the D/H value.
This suggests that the D/H ratio in 67P’s coma remains constant throughout the entire Rosetta mission phase, covering one third of 67P's orbit. Additionally, considering the large diversity of the conditions under which the data have been observed, the D/H ratio in 67P's coma seems to be independent of heliocentric distance, level of cometary activity, observational phase angle as well as sub-S/C latitudes. The D/H ratio did not even significantly change during extreme situations such as a maximally active southern hemisphere or a phase angle of almost 70\textdegree. Figure~\ref{fig:H2OTime} shows the individual D/H ratio data points alongside their corresponding H\textsubscript{2}O gas production rate \citep{Laeuter2020}, phase angle, latitude and cometary distance to the Sun. For the H\textsubscript{2}O gas production, \citet{Laeuter2020} reported minimum and maximum values according to their uncertainty estimation. No H\textsubscript{2}O gas production values were reported by these authors for the time between 2016-03-13 and the end of the measurements during the second equinox. \citet{Combi2020} provided gas production values for individual measurement points acquired with a different approach, with their results for the overall variation of the H\textsubscript{2}O gas production rate being in reasonable agreement with \citet{Laeuter2020}.

For the absolute value, the systematic uncertainty is added. 
This systematic uncertainty affects all data points equally and leads to an absolute mean D/H ratio of $(5.01 \pm 0.40)\times10^{-4}$.
This is consistent with the previously published values of $(5.3 \pm 0.7)\times10^{-4}$ found by \cite{Altwegg2015, Altwegg2017}. These earlier values were determined before a better understanding of the behaviour of the pixel gain and the overall gain of the DFMS over time was available \citep{DeKeyser2019, Schroeder2019b}. By extending the number of spectra from 26 and 18 in \citet{Altwegg2015} and \citet{Altwegg2017}, respectively, to 150 spectra in this work, and due to the better characterization of the DFMS over time, we were able to improve on the uncertainty. For statistical reasons, this uncertainty is inversely proportional to the square root of the number of spectra and thus greatly decreased by the large number of spectra considered here.

\subsection{\textsuperscript{16}O/\textsuperscript{17}O}
In addition to the signature of HDO, H\textsubscript{2}\textsuperscript{17}O has been measured on \emph{m/z}~19. Together with the already examined H\textsubscript{2}\textsuperscript{16}O on \emph{m/z}~18, the isotopic ratio of \textsuperscript{16}O/\textsuperscript{17}O could be derived. This has already been done by \citet{Schroeder2019b} in a "Note added in proof". However, the authors only investigated data from two distinct dates and only used 35 spectra. 
With the 150 spectra investigated for the time periods given in Table~\ref{table:H2OSummary}, an updated value can now be presented. Table~\ref{table:OxygenSummary} shows the relative mean \textsuperscript{16}O/\textsuperscript{17}O ratios for the different mission phases, considering only statistical and fit uncertainties. As with section \ref{HDO/H2O}, the mean values are weighted means and the uncertainty is inversely proportional to the square root of the number of spectra. The relative overall mean value over all evaluated spectra was found to be $2347 \pm 53$. All periods are consistent with this mean value within the 1$\sigma$ uncertainty of 2.3\%. There is no observable trend in the \textsuperscript{16}O/\textsuperscript{17}O ratio between the periods considered. This is in line with the invariability of the \textsuperscript{16}O/\textsuperscript{18}O ratio in \citet{Schroeder2019b}. It is, however, in contrast with their average values for the \textsuperscript{16}O/\textsuperscript{17}O ratios, as their \textsuperscript{16}O/\textsuperscript{17}O ratio for the first date is approximately 40\% higher than the \textsuperscript{16}O/\textsuperscript{17}O ratio for the second date. An explanation for this might be that \citet{Schroeder2019b} did not include H\textsubscript{3}\textsuperscript{16}O in their evaluation of \emph{m/z}~19 spectra. However, all three molecules, H\textsubscript{2}\textsuperscript{17}O, HDO and H\textsubscript{3}\textsuperscript{16}O, need to be included in the analysis as their peaks overlap significantly and the influence of H\textsubscript{3}\textsuperscript{16}O should not be ignored. 
Including the systematic uncertainties of the gains affecting all data points equally, gives an absolute mean \textsuperscript{16}O/\textsuperscript{17}O ratio of $2347 \pm 191$. This represents an approximately 11\% enrichment of \textsuperscript{17}O compared to the value for terrestrial water of $(2632 \pm 69)$ \citep{Meija2016} and is in line with the enrichment of \textsuperscript{18}O in 67P's coma \citep{Schroeder2019b}. The \textsuperscript{16}O/\textsuperscript{17}O ratio we derived is compatible within uncertainties with the value reported in the "Note added in proof" in \citet{Schroeder2019b}.

\subsection{Linear alkanes}
The D/H and \textsuperscript{13}C/\textsuperscript{12}C ratios of the first four linear alkanes, namely, methane, ethane, propane and butane, have been evaluated. For all alkanes, $\mathrm{C\textsubscript{n}H\textsubscript{y}}$, taking into account the statistical correction for the different possible positions of the rare isotopes in the molecule, the D/H and  \textsuperscript{13}C/\textsuperscript{12}C ratios are obtained by dividing the measured abundance ratios by $\mathrm{1/y}$ and $\mathrm{1/n}$, respectively. The alkanes were not always at a detectable level over the entire mission. This required an individual selection of suitable time periods for each of the molecules. For each hydrocarbon, the results will be presented separately in the following subsections.

\begin{table}
\caption{\textsuperscript{16}O/\textsuperscript{17}O in water during different mission phases.}             
\label{table:OxygenSummary}      
\centering          
\begin{tabular}{l l }     
\hline
\noalign{\smallskip}
Mission Phase & \textsuperscript{16}O/\textsuperscript{17}O\\
\noalign{\smallskip}
\hline
\noalign{\smallskip}
1st Equinox & 2317 ± 91\\
Perihelion & 2318 ± 115\\
Peak gas production & 2398 ± 141\\
2nd Equinox & 2379 ± 97\\
\noalign{\smallskip}
\hline
\noalign{\smallskip}
Relative mean ratio & 2347 ± 53\\
Absolute mean ratio & 2347 ± 191\\
\noalign{\smallskip}
\hline                    
\end{tabular}
\end{table}

\subsubsection{Methane (CH\textsubscript{4})}
The methane (CH\textsubscript{4}) signature was observed clearly from mid-August 2016 until the beginning of September 2016. Hence, 12 spectra from this time period have been evaluated. Spectra with \emph{m/z} 16 and \emph{m/z} 17 have been investigated. Sample spectra are shown in Fig.~\ref{fig:sampleSpectraMethane}. The \emph{m/z} 16 spectra used gain step 15, whereas the \emph{m/z} 17 spectra used gain step 16. Thus, a gain step correction was needed. The gain step corrections were calibrated with data acquired shortly before this period \citep{Schroeder2019a}. Gain step 16 was used as the baseline by \citet{Schroeder2019a} for the gain step corrections. Consequently, the gain step correction was simple for the ratio calculated from the data considered here. On \emph{m/z} 17, \textsuperscript{13}CH\textsubscript{4} and CH\textsubscript{3}D are slightly over-lapping and a clear distinction is not always straightforward. This additional uncertainty has been included in the overall uncertainty.

The \textsuperscript{13}C/\textsuperscript{12}C ratio has already been evaluated several times for 67P by \citet{Haessig2017} and \citet{Rubin2017} for carbon dioxide (CO\textsubscript{2}), carbon monoxide (CO), ethylene (C\textsubscript{2}H\textsubscript{4}) and the ethyl radical C\textsubscript{2}H\textsubscript{5} and has been shown to be independent of the parent molecule. Hence, the value of \textsuperscript{13}C/\textsuperscript{12}C = $(1.19 \pm 0.06)\times10^{-2}$ derived from CO\textsubscript{2} by \citet{Haessig2017} will be used as a comparison for the values derived in this work.

From the measurements of CH\textsubscript{4}, \textsuperscript{13}CH\textsubscript{4} and CH\textsubscript{3}D, the D/H and \textsuperscript{13}C/\textsuperscript{12}C ratios could be derived taking into account the statistical correction for the four possible positions the D atom can take in the molecule. An average value of D/H = $(2.41 \pm 0.29)\times10^{-3}$ is found in methane (Fig.~\ref{fig:AlkanesSummary} and Table~\ref{table:AlkanesSummary}). This is 4.8 times larger than the D/H ratio from HDO/H\textsubscript{2}O but 7.5 times smaller than the D/H ratio from D\textsubscript{2}O/HDO \citep{Altwegg2017}. The corresponding ratio of \textsuperscript{13}C/\textsuperscript{12}C is $(1.14 \pm 0.13)\times10^{-2}$, which is consistent with \citet{Haessig2017}. This is additional evidence of the \textsuperscript{13}C/\textsuperscript{12}C ratio being independent of the parent molecule.

\subsubsection{Ethane (C\textsubscript{2}H\textsubscript{6})}
\begin{figure}
\centering
\includegraphics[width=\hsize]{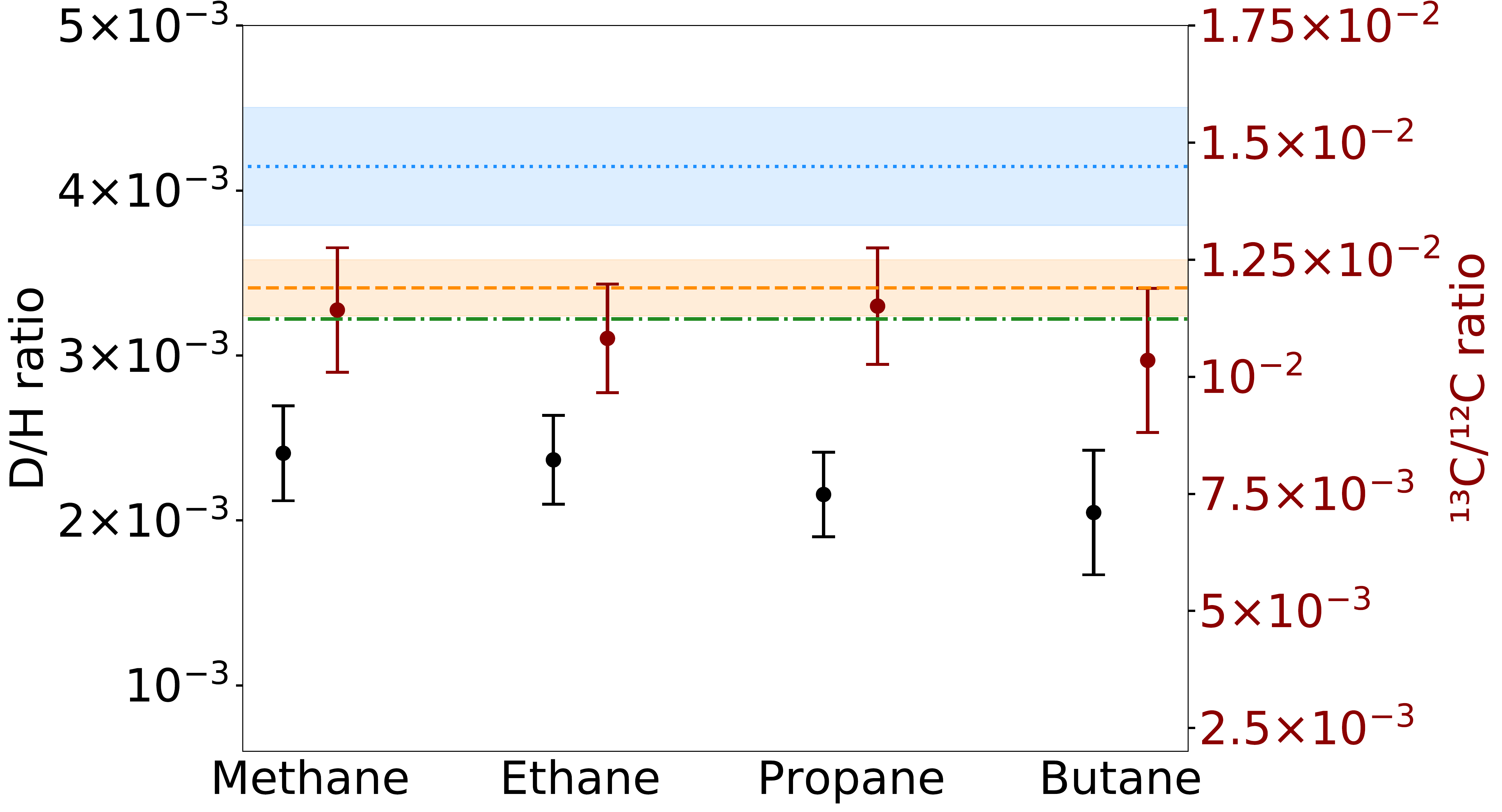}
\caption{D/H (black) and \textsuperscript{13}C/\textsuperscript{12}C (red) ratios of the first four linear alkanes compared with \textsuperscript{13}C/\textsuperscript{12}C values in 67P's CO\textsubscript{2} (\citealp{Haessig2017}, orange), in the Earth (\citealp{Wilson1999}, green) and in the local ISM (\citealp{Wilson1999}, blue).}
\label{fig:AlkanesSummary}
\end{figure}

The evaluation of ethane (C\textsubscript{2}H\textsubscript{6}) was very similar to the one for methane. 20 spectra with \emph{m/z} 30 and \emph{m/z} 31 have been evaluated for time periods in the beginning of October 2014 and during the second equinox in March 2016. Here, the gain steps were the same on both \emph{m/z} spectra and no gain correction was needed. Similar to methane, an overlap between \textsuperscript{13}C\textsuperscript{12}CH\textsubscript{6} and C\textsubscript{2}H\textsubscript{5}D appears on \emph{m/z} 31. Again, this uncertainty has been included in the overall uncertainty of the corresponding ratios. Sample spectra for \emph{m/z} 30 and \emph{m/z} 31 are shown in Fig.~\ref{fig:sampleSpectraAlkanes}. NO and C\textsuperscript{18}O strongly overlap and cannot be clearly separated. For this reason, they are fitted together as one peak.

Accounting for the statistical correction for the different possible positions of the rare isotopes in the molecule, an average D/H ratio of $(2.37 \pm 0.27)\times10^{-3}$ (Fig.~\ref{fig:AlkanesSummary} and Table~\ref{table:AlkanesSummary}) and a \textsuperscript{13}C/\textsuperscript{12}C ratio of $(1.08 \pm 0.12)\times10^{-2}$ have been obtained. This is consistent, within uncertainties, with \citet{Haessig2017} as well as the result for methane.

\begin{table}
\caption{D/H and \textsuperscript{13}C/\textsuperscript{12}C in linear alkanes.}             
\label{table:AlkanesSummary}      
\centering          
\begin{tabular}{c c c }     
\hline
\noalign{\smallskip}
Alkane & D/H & \textsuperscript{13}C/\textsuperscript{12}C\\
\noalign{\smallskip}
\hline
\noalign{\smallskip}
Methane & (2.41 ± 0.29) · 10\textsuperscript{-3} & (1.14 ± 0.13) · 10\textsuperscript{-2}\\
Ethane & (2.37 ± 0.27) · 10\textsuperscript{-3} & (1.08 ± 0.12) · 10\textsuperscript{-2}\\
Propane & (2.16 ± 0.26) · 10\textsuperscript{-3} & (1.15 ± 0.12) · 10\textsuperscript{-2}\\
Butane & (2.05 ± 0.38) · 10\textsuperscript{-3} & (1.04 ± 0.15) · 10\textsuperscript{-2}\\
\noalign{\smallskip}
\hline                    
\end{tabular}
\end{table}

\subsubsection{Propane (C\textsubscript{3}H\textsubscript{8})}
14 Spectra with \emph{m/z} 44 and \emph{m/z} 45 from the second equinox in March 2016 have been evaluated for propane (C\textsubscript{3}H\textsubscript{8}). Sample spectra are shown in Fig.~\ref{fig:sampleSpectraPropane}. The spectra with \emph{m/z} 44 contain a very large amount of CO\textsubscript{2}. Consequently, a small gain step was automatically selected by the DFMS while acquiring these spectra. The spectra measured around \emph{m/z} 45 on the other hand showed consistently lower count rates and were thus measured on a larger gain step. For this reason, a gain correction needed to be applied before the spectra could be compared. The gain steps differed by up to four gain steps as some spectra of \emph{m/z} 44 were acquired with a very low gain step (i.e. gain step 11). Low gain steps were difficult to calibrate during the calibration measurements and complicate the gain step corrections. However, \citet{Haessig2017} obtained the \textsuperscript{13}C/\textsuperscript{12}C ratio in CO\textsubscript{2} at times when the gain steps of \emph{m/z} 44 and 45 were much closer. From the \textsuperscript{13}C/\textsuperscript{12}C ratio in CO\textsubscript{2} we thus inferred a gain correction for our measurements.

After applying the gain correction and accounting for the different possible positions of the rare isotopes in the molecule, an average D/H ratio of $(2.16 \pm 0.26)\times10^{-3}$ was found for propane (C\textsubscript{3}H\textsubscript{8}, Fig.~\ref{fig:AlkanesSummary} and Table~\ref{table:AlkanesSummary}). For \textsuperscript{13}C/\textsuperscript{12}C from \textsuperscript{13}C\textsuperscript{12}C\textsubscript{2}H\textsubscript{8} and C\textsubscript{3}H\textsubscript{8}, the value is $(1.15 \pm 0.12)\times10^{-2}$. Again, this value is compatible with the value from \citet{Haessig2017} and the other linear alkanes.

\subsubsection{Butane (C\textsubscript{4}H\textsubscript{10})}
For butane (C\textsubscript{4}H\textsubscript{10}), 11 spectra with \emph{m/z} 58 and \emph{m/z} 59 have been evaluated from data acquired during the second equinox in March 2016. Here, a gain correction was unnecessary as both, \emph{m/z} 58 and \emph{m/z} 59, were measured with the highest gain available. Sample spectra for butane are shown in Fig.~\ref{fig:sampleSpectraButane}. 

Taking into account the different possible positions of the D or \textsuperscript{13}C in the molecule, butane (C\textsubscript{4}H\textsubscript{10}) showed a D/H ratio of $(2.05 \pm 0.38)\times10^{-3}$ and a \textsuperscript{13}C/\textsuperscript{12}C ratio from \textsuperscript{13}C\textsuperscript{12}C\textsubscript{3}H\textsubscript{10} and C\textsubscript{4}H\textsubscript{10} of $(1.04 \pm 0.15)\times10^{-2}$ (Fig.~\ref{fig:AlkanesSummary} and Table~\ref{table:AlkanesSummary}). As with all of the other linear alkanes considered above, the \textsuperscript{13}C/\textsuperscript{12}C ratio is consistent with \citet{Haessig2017} and the other linear alkanes.
\section{Discussion}
The ROSINA/DFMS measurements show that the D/H ratio in water does not change during 67P’s passage around the Sun between May 2015 and March 2016. It is clear, that the instrument's observations represent an average of the illuminated surface, even though they have been measured at different positions. Hence, we cannot examine any point-to-point variability on the surface itself. However, given the large variability of the phase angles and sub-S/C latitudes during the evaluated measurement phases and their association with different spacecraft distances to the comet, we can conclude that the D/H ratio in water in 67P's coma is independent of heliocentric distance, level of cometary activity, Rosetta's phase angle as well as sub-S/C latitude (Fig.~\ref{fig:H2OTime}). 
The relative overall mean value, considering only statistical and fit uncertainties, has a 1$\sigma$ variation of 2.0\% with all investigated periods being consistent.
The derived D/H ratio for water is compatible with values previously published in \cite{Altwegg2015, Altwegg2017}. However, the new values presented in this work are based on a larger number of measurements and hence have smaller error margins. The most accurate absolute value for D/H in HDO/H\textsubscript{2}O we obtained from our data is $(5.01 \pm 0.40)\times10^{-4}$, where the uncertainty includes all statistical and systematic uncertainties.

\citet{Paganini2017} and \citet{Biver2016} reported different values for the D/H ratio in water for comet Lovejoy pre- and post-perihelion. \citet{Paganini2017} favoured the explanation of a systematic difference between the two observations by \citet{Biver2016} as a reason for the changing D/H ratio observed for comet Lovejoy. If the results for comet 67P are valid for other comets, our study indicates a constant D/H ratio, within uncertainties, and therefore supports the hypothesis of a systematic difference rather than a change in the D/H ratio of comet Lovejoy.

\citet{Lis2019} proposed that the D/H ratio in cometary water correlates with the nucleus’ active area fraction. \citet{Fulle2021} modelled this scenario and suggested that the fraction of water-rich and water-poor pebbles influences the D/H ratio in the comet's coma. The data evaluated for this paper show that the D/H ratio is independent of 67P's activity (in the form of H\textsubscript{2}O outgassing) and Rosetta's relative position in terms of phase angle and sub-spacecraft latitude and hence, do not show any signs of such a scenario for 67P. 

\begin{figure*}
\centering
\includegraphics[width=\hsize]{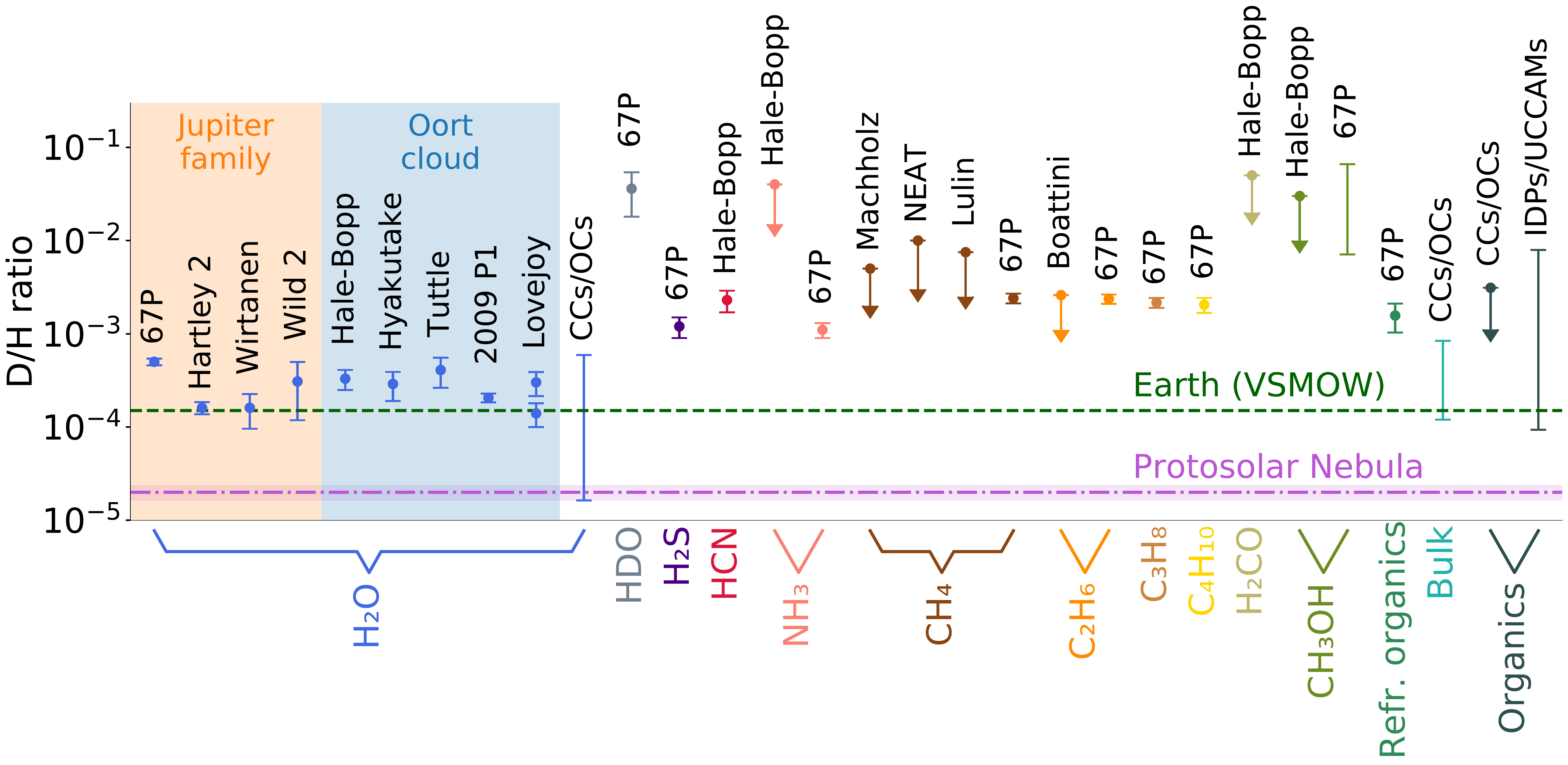}
\caption{D/H ratios of water and organic molecules measured in different comets compared to values from the Protosolar Nebula (purple line, \citealp{Geiss2003}), the Earth (green line, \citealp{Wilson1999}), carbonaceous chondrites (CC), ordinary chondrites (OC), interplanetary dust particles (IDP) and ultracarbonaceous Antarctic micrometeorites (UCCAM). D/H in HDO is equal to 2 · D\textsubscript{2}O/HDO. Full references are given in Table~\ref{table:LiteratureD}.}
\label{fig:Organics}
\end{figure*}

Measurements taken of the first four linear alkanes in 67P's coma show that their D/H ratios are all consistent within uncertainties. The derived values are larger than the aforementioned ratio obtained from HDO/H\textsubscript{2}O by a factor of 4.1 to 4.8, but smaller than the D/H ratio obtained from D\textsubscript{2}O/HDO \citep{Altwegg2017}.
In addition, the D/H ratio in the alkanes is slightly larger than the ratio obtained from HDS \citep{Altwegg2017} but still of the same order of magnitude.
The first-time detection of mono- and di-deuterated methanol in a cometary coma was published by \citet{Drozdovskaya2021}. The authors evaluated Rosetta/ROSINA data for 67P. With the ROSINA instruments, it is not possible to distinguish between the different chemical compositions of D-methanol (CH\textsubscript{3}OD and CH\textsubscript{2}DOH) and D\textsubscript{2}-methanol (CH\textsubscript{2}DOD and CHD\textsubscript{2}OH), respectively. Moreover, different approaches for the calculation of the D/H ratio in methanol are possible and it cannot be judged which of the pathways is more reliable. Consequently, although it was not possible to deduce a single D/H ratio in CH\textsubscript{3}OH, a range of 0.71-6.6 per cent was given by the authors. This accounts for the different isomers of methanol and includes statistical error propagation in the ROSINA measurements. The authors propose that methanol and its deuterated isotopologues in comet 67P must have formed in the prestellar core that preceded our Solar System and at a time when it was at a temperature of 10–20~K. Moreover, it is assumed that methanol is a pivotal precursor to complex organic molecules, and hence, could be a source of deuterium for such species \citep{Oba2016}.
The presented D/H value in methanol is much larger than the ratios obtained for the first four linear alkanes. However, \citet{Drozdovskaya2021} demonstrate that the upper boundary of 6.6\% of their determined D/H range would only apply in the extreme case where all D-methanol was in the form of CH\textsubscript{3}OD. In the much more likely case that D-methanol exists in the form of different isomers \citep{Ratajczak2011}, the D/H in methanol would be lower and thus comparable to the D\textsubscript{2}O/HDO ratio of (1.8 ± 0.9) · 10\textsuperscript{-2} from \citet{Altwegg2017}.

\citet{Furuya2016} describe the development of ice structures during the formation of protostellar cores with two layers from molecular clouds. The first layer is the main formation stage of H\textsubscript{2}O ice. The second, outer layer is CO/CH\textsubscript{3}OH-rich and includes material that underwent enhanced deuteration processes due to low temperatures (T < 20 K). Their model shows higher levels of deuterium fractionation of formaldehyde and methanol from the outer layer than in water in the inner layer and gives similar D/H ratios for methanol and D\textsubscript{2}O. They suggest that this difference reflects the epochs of the molecules' formation as water ice is formed at an earlier stage of protostellar cloud condensation at elevated temperatures than the ices of formaldehyde and methanol. These conclusions might explain why at 67P the D/H ratios for D\textsubscript{2}O/HDO and methanol are similar but much larger compared to D/H in H\textsubscript{2}O. How such a scenario affects the deuteration in hydrocarbons, however, requires further investigation.

Measurements of organics in other comets, for instance the D/H ratio in HCN in comet C/1995 O1 (Hale-Bopp) \citep{Crovisier2004}, match the values for the first four linear alkanes within uncertainties.

\citet{Paquette2021} presented the first in-situ measurements of the D/H ratios in organic refractory components of cometary dust particles. These cometary dust particles have been captured on metal targets within the coma of comet 67P. The particles were then imaged by a microscope camera and a fraction of them were analysed with the Cometary Secondary Ion Mass Analyzer (COSIMA), a time-of-flight secondary ion mass spectrometer \citep{Kissel2007}. The incident velocities of the particles COSIMA collected were low and they did not suffer a large degree of thermal alteration. Larger thermal alterations occur in flyby missions, where incident velocities experienced by particles are orders of magnitude larger \citep{Paquette2021}.
The D/H ratio of $(1.57 \pm 0.54)\times10^{-3}$ in the organic refractory components of 67P's cometary dust is comparable to our D/H ratios in linear alkanes. It is thus also about an order of magnitude higher than the VSMOW for the D/H ratio on Earth. \citet{Paquette2021} state, that this relatively high value puts forward the theory that refractory carbonaceous matter in comet 67P is less processed than the most primitive insoluble organic matter (IOM) in meteorites. 

\citet{Bonev2009} reported an upper limit of $5\times10^{-3}$ for the D/H ratio in methane in comet C/2004 Q2 (Machholz), while \citet{Kawakita2005} determined an upper limit for the D/H ratio of $1\times10^{-2}$ for comet C/2001 Q4 (NEAT), and \cite{Gibb2012} found an upper limit of $7.5\times10^{-3}$ for comet C/2007 N3 (Lulin). The D/H ratio we determined in methane for 67P is about a factor of two lower than the smallest previously obtained upper limit for this molecule.

An upper limit for the D/H ratio in ethane of $2.6\times10^{-3}$ from modelled emission spectra of comet C/2007 W1 (Boattini) has been determined by \citet{Doney2020}. Hence, for ethane, our D/H ratio for 67P is comparable to this upper limit.

This work is the first to present an isotopic ratio for methane, ethane, propane and butane for comets and no other values are available for comparison.

A comparison of the D/H ratios investigated here with values obtained from different comets and on different organic molecules is shown in Fig.~\ref{fig:Organics}. D/H ratios from the Protosolar Nebula, Earth, carbonaceous chondrites (CC), ordinary chondrites (OC), interplanetary dust particles (IDP) and ultracarbonaceous Antarctic micrometeorites (UCCAM) are added for comparison.
The D/H ratio from HDO/H\textsubscript{2}O is larger for most of the observed comets compared to the terrestrial value, though they show large variations. Variations are also observed within the comet families, the JFCs and the Oort cloud comets (OCC). It also seems that organic compounds in the comets investigated exhibit even larger D/H ratios than water. A comparison of the D/H ratios derived from cometary organics, chondrites, and IDPs to values from the Protosolar Nebula and the VSMOW reveals a pronounced deuterium enrichment in Solar System objects in general. \citet{Hoppe2018} suggested that 67P might be particularly primordial and might have conserved large amounts of presolar matter due to the fact that its D/H ratio corresponds to the highest values proposed for comets to date.
Water in chondrites has D/H ratios in between those of the Protosolar Nebula and the highest cometary values. On the other hand, chondritic IOM shows strong D-enrichment as compared to VSMOW. According to \citet{Alexander2010}, this deuterium enrichment is not a signature of the primordial H in the presolar cloud, but is caused by different processes. Moreover, \citet{Duprat2010} analysed ultracarbonaceous micrometeorites recovered from central Antarctic snow and found extreme deuterium enrichment in large areas of the organic matter contained therein. In addition, crystalline minerals embedded in the micrometeoritic organic matter have been identified. According to the authors, this suggests that this organic matter reservoir may have formed within the Solar System itself and was not inherited from presolar times. As a summary of their findings, the high D/H ratios, the high organic matter content, and the associated minerals are said to favour an origin from the cold regions of the protoplanetary disc \citep{Duprat2010}.

\begin{figure}
\centering
\includegraphics[width=\hsize]{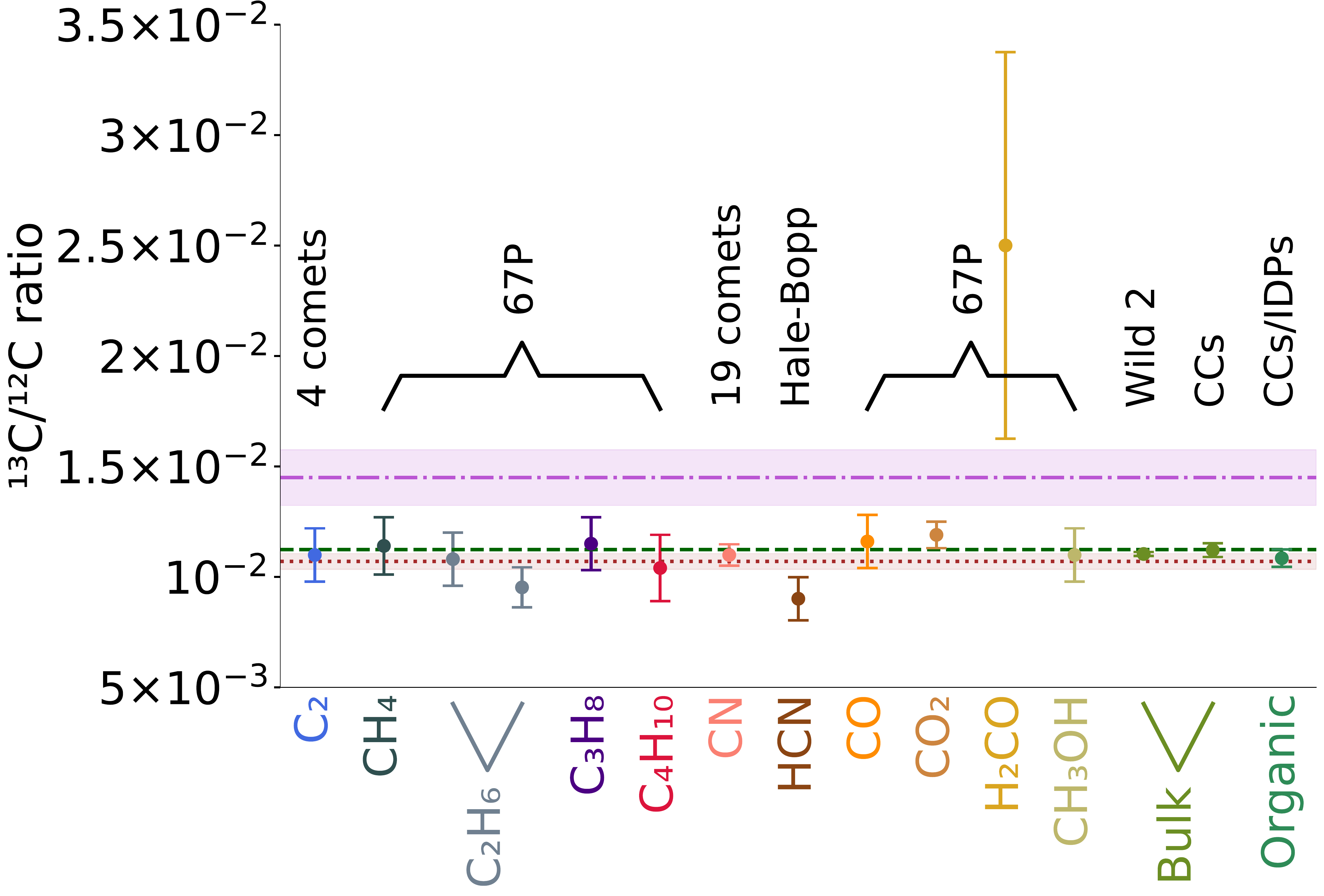}
\caption{\textsuperscript{13}C/\textsuperscript{12}C ratios of different organic molecules measured in different comets, carbonaceous chondrites (CC) and interplanetary dust particles (IDP) compared to the terrestrial value (green line, \citealp{Lyons2018}), the Sun (brown line, \citealp{Lyons2018}) and the ISM (purple line, \citealp{Wilson1999}). For C\textsubscript{2} and CN, data from 4 and 19 comets, respectively, have been considered. Full references are given in Table~\ref{table:LiteratureC}.}
\label{fig:OrganicsCarbon}
\end{figure}

According to \citet{Cleeves2016}, the D/H ratio in both water and organics can become chemically enhanced in cold environments exposed to ionizing radiation. The authors proposed the cold interstellar medium, activated by galactic cosmic rays, and the outermost regions of the protoplanetary disc in the presence of stellar or non-stellar ionization, as two possible environments where this deuterium enrichment could occur. In an earlier study, \citet{Cleeves2014} state that a considerable fraction of the Solar System’s water predates the Sun and that a certain amount of such interstellar ice survived the formation of the Solar System and has been incorporated into planetesimals. The authors also identified two factors which might lead to the even higher degree of deuterium enrichment in protoplanetary disc organics as compared to water: (1) the higher volatility and abundance of CO, which is the main carbon reservoir, as compared to O (atomic oxygen), which is the main precursor for water formation, and, (2) a more favourable chemistry for deuterium-fractionation in organics than in water due to a higher exothermicity in the chemical formation reaction \citep{Cleeves2016}.

Embedded protostars in low-mass star-forming regions exhibit D/H ratios in their water similar to values found in comets \citep{Persson2014}. On the other hand, isolated protostars have D/H ratios of more than double the values observed in embedded protostars \citep{Jensen2019}, and their D/H ratios are thus more similar to those of cometary organics.
The high D/H ratios in cometary organic compounds in general, suggest that these species may be inherited from the presolar molecular cloud from which the Solar System formed.

The alkanes investigated show \textsuperscript{13}C/\textsuperscript{12}C ratios compatible with published values for CO \citep{Rubin2017} and CO\textsubscript{2} \citep{Haessig2017} in 67P and the \textsuperscript{13}C/\textsuperscript{12}C ratio in the Solar System \citep{Wilson1999}. \citet{Altwegg2020} found a \textsuperscript{13}C/\textsuperscript{12}C ratio in ethane of $(0.95 \pm 0.1)\times10^{-2}$ which matches the results presented in this work within uncertainties. These authors also revealed that the \textsuperscript{13}C/\textsuperscript{12}C ratio varies for different molecules in 67P's coma, but that, except for H\textsubscript{2}CO with its large uncertainty, the \textsuperscript{13}C/\textsuperscript{12}C ratios are in the same range as our values. This picture is supported by data from other comets and even bulk and organic CCs, where the \textsuperscript{13}C/\textsuperscript{12}C ratios for different molecules are similar \citep{Bockelee2015, Hoppe2018}. A comparison of \textsuperscript{13}C/\textsuperscript{12}C ratios for different organic molecules measured in different comets and other Solar System objects is shown in Fig.~\ref{fig:OrganicsCarbon}. All these values lie below the local ISM value \citep{Wilson1999} but are mostly compatible with the terrestrial and the solar value \citep{Lyons2018}. This indicates that isotopic fractionation may have occurred over time and was, at least for most organic molecules, independent of the molecular structure.

\begin{figure}
\centering
\includegraphics[width=\hsize]{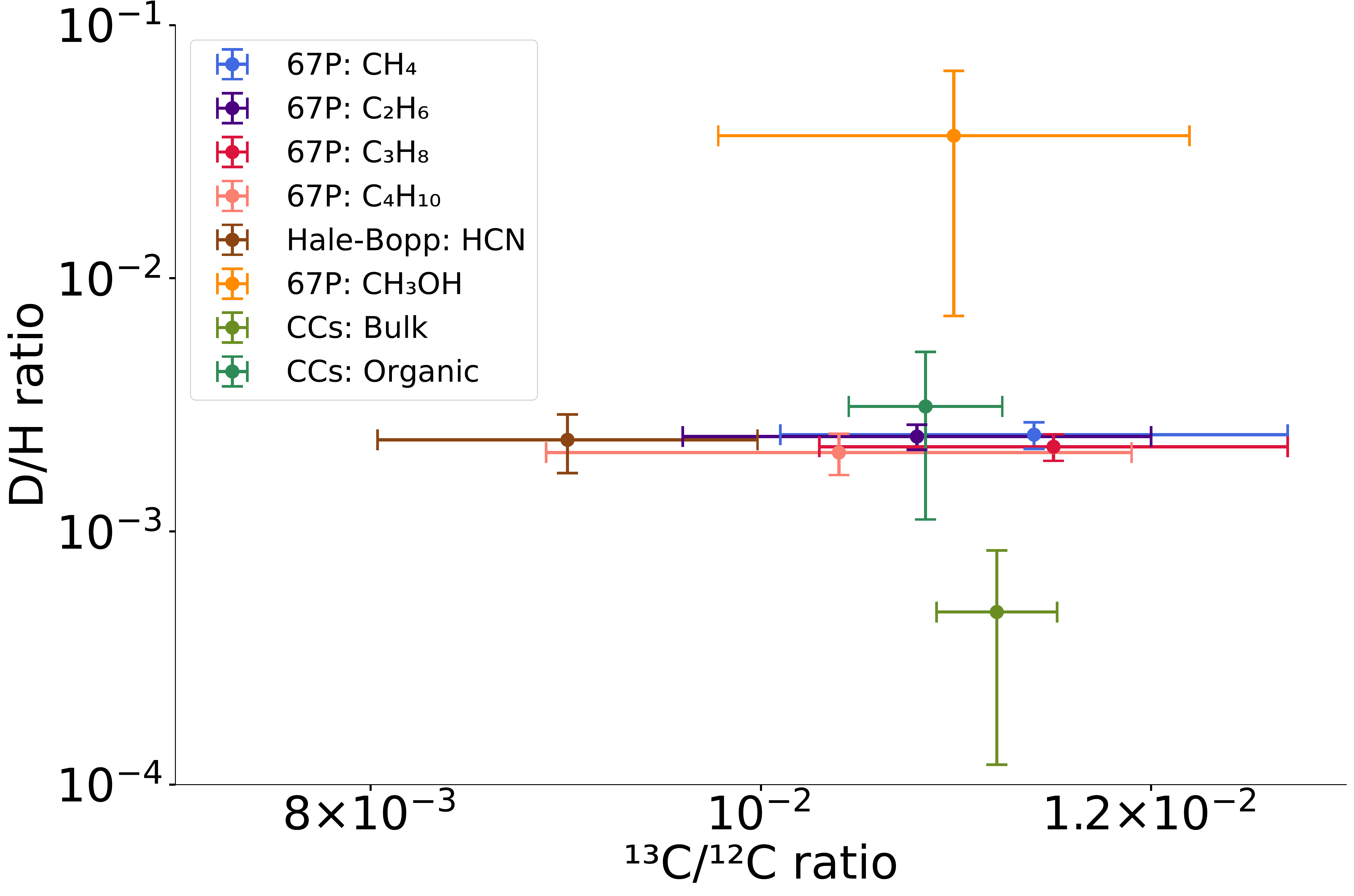}
\caption{D/H ratios compared to \textsuperscript{13}C/\textsuperscript{12}C ratios of different organic molecules measured in comets 67P and Hale-Bopp and in bulk and organic carbonaceous chondrites (CC). Full references are given in Table~\ref{table:LiteratureD} for the D/H ratios and in Table~\ref{table:LiteratureC} for the \textsuperscript{13}C/\textsuperscript{12}C ratios.}
\label{fig:OrganicsDC}
\end{figure}

In conjunction with the small variations in the \textsuperscript{13}C/\textsuperscript{12}C ratios and the large variations in the D/H ratios, Fig.~\ref{fig:OrganicsDC} illustrates that there is no correlation between the \textsuperscript{13}C/\textsuperscript{12}C ratio and the D/H ratio for comets and CCs.

\section{Summary \& conclusions}
In this work, we investigated the isotopic ratios of water and the four simplest alkanes in comet 67P's inner coma. The most relevant findings can be summarized as follows:
\begin{itemize}
    \item The D/H ratio in water in 67P's coma, measured with ROSINA/DFMS, is independent of the heliocentric distance, the level of cometary activity, the spacecraft's phase angle and the sub-spacecraft latitude.
    \item A 1$\sigma$ variation of 2.0\% is included in the relative overall mean value. All values derived from the investigated periods are consistent with this value.
    \item From our data, we obtained an absolute D/H ratio from HDO/H\textsubscript{2}O of $(5.01 \pm 0.40)\times10^{-4}$. Many comets exhibit larger D/H ratios in water as compared to the terrestrial value. However, both cometary families, JFCs and OCCs, also include comets with values comparable to the VSMOW value. Hence, the implications for cometary contributions to terrestrial water remain unclear if only cometary water is examined.
    \item The \textsuperscript{16}O/\textsuperscript{17}O ratio in water in 67P's coma was determined to be constant throughout the mission, with a relative 1$\sigma$ variation of 2.3\%. An absolute \textsuperscript{16}O/\textsuperscript{17}O ratio of $2347 \pm 191$ has been found.
    \item The four simplest linear alkanes show larger D/H ratios than 67P's water by a factor of 4.1 to 4.8. Their D/H ratio values are consistent with data from other organic molecules and from different comets.
    \item A comparison between different sources of cometary matter showed that organic molecules generally exhibit higher D/H ratios than water for all comets reviewed in this work.
    \item No correlation was found between the \textsuperscript{13}C/\textsuperscript{12}C ratio and the D/H ratio for different cometary molecules.
\end{itemize}

The observed invariability of the D/H ratio in 67P's coma opposes theories of a non-steady-state regime of water ice sublimation occurring in sporadic time intervals along the comet's orbit. However, this invariability needs to be confirmed for other comets with further measurements and with other measurement approaches. Additionally, 67P's close apparition in November 2021 has been an excellent opportunity to re-measure the D/H ratio using spectroscopic approaches and upcoming results are highly anticipated.
On the other hand, to further constrain the history and origin of organic matter in the Solar System, more cometary and other Solar System objects' data need to be analysed and more studies are required to investigate these species' formation pathways.
\begin{acknowledgements}
We thank two anonymous referees for their constructive feedback that helped to improve the paper.
Work at the University of Bern was funded by the State of Bern and the Swiss National Science Foundation (200020\_182418). S.F.W. acknowledges the financial support of the SNSF Eccellenza Professorial Fellowship (PCEFP2\_181150). The results from ROSINA would not be possible without the work of the many engineers, technicians, and scientists involved in the mission, in the Rosetta spacecraft, and in the ROSINA instrument team over the past 20 years, whose contributions are gratefully acknowledged. Rosetta is a European Space Agency (ESA) mission with contributions from its member states and NASA. We thank herewith the work of the whole ESA Rosetta team. All ROSINA flight data have been released to the PSA archive of ESA and to the PDS archive of NASA.

The data used in this article are available in the European Space Agency's Planetary Science Archive (PSA), at archives.esac.esa.int/psa/

\end{acknowledgements}

~~\newline

\bibliographystyle{aa} 
\bibliography{literature}

\onecolumn
\begin{appendix}
\section{Mass spectra showing signatures of methane, propane and butane and their isotopologues}

\begin{figure}[h]
    \centering
    \includegraphics[width=.5\textwidth]{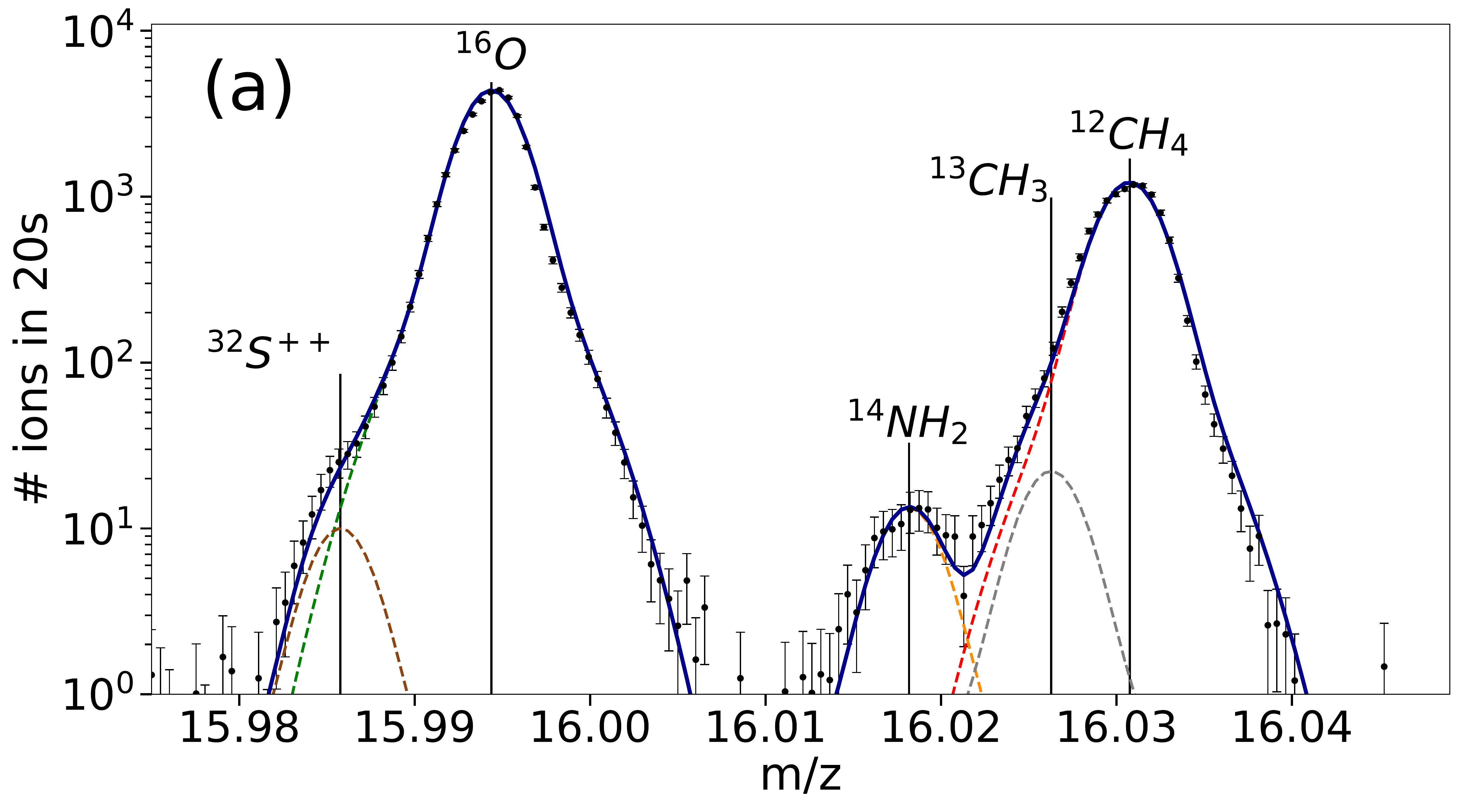}\hfill
    \includegraphics[width=.5\textwidth]{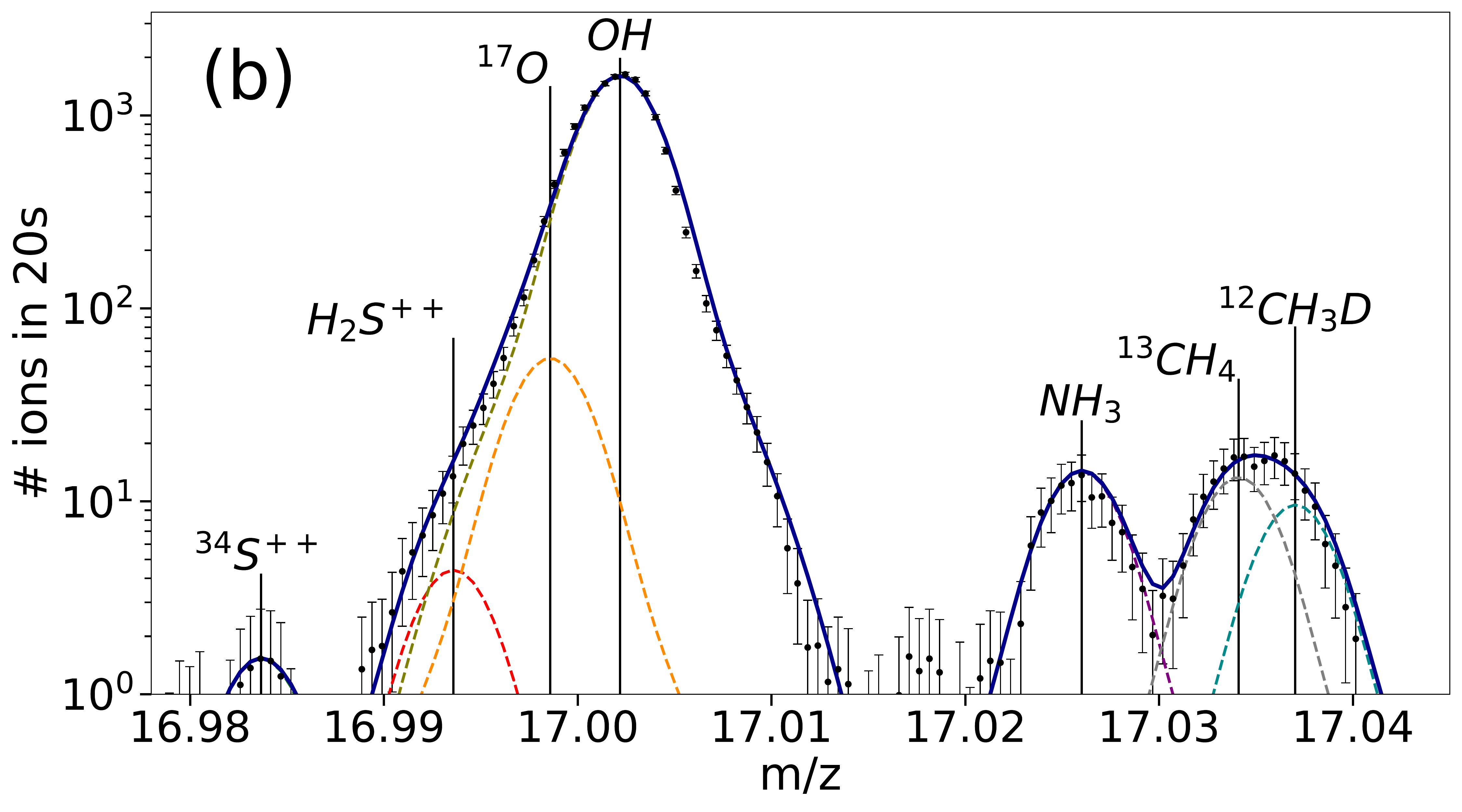}
    \caption{Sample mass spectra for \emph{m/z} 16 and 17 showing the signatures of the isotopologues of methane. \textit{Panel a}: \emph{m/z} 16 from 2016-03-09 11:13 (UTC). \textit{Panel b}: \emph{m/z} 17 from 2016-03-09 11:14 (UTC). Measured data are represented by black dots including their statistical uncertainties. The individual mass fits and the total sum of the fits are shown with coloured lines.}
    \label{fig:sampleSpectraMethane}
\end{figure}

\begin{figure}[h]
    \centering
    \includegraphics[width=.5\textwidth]{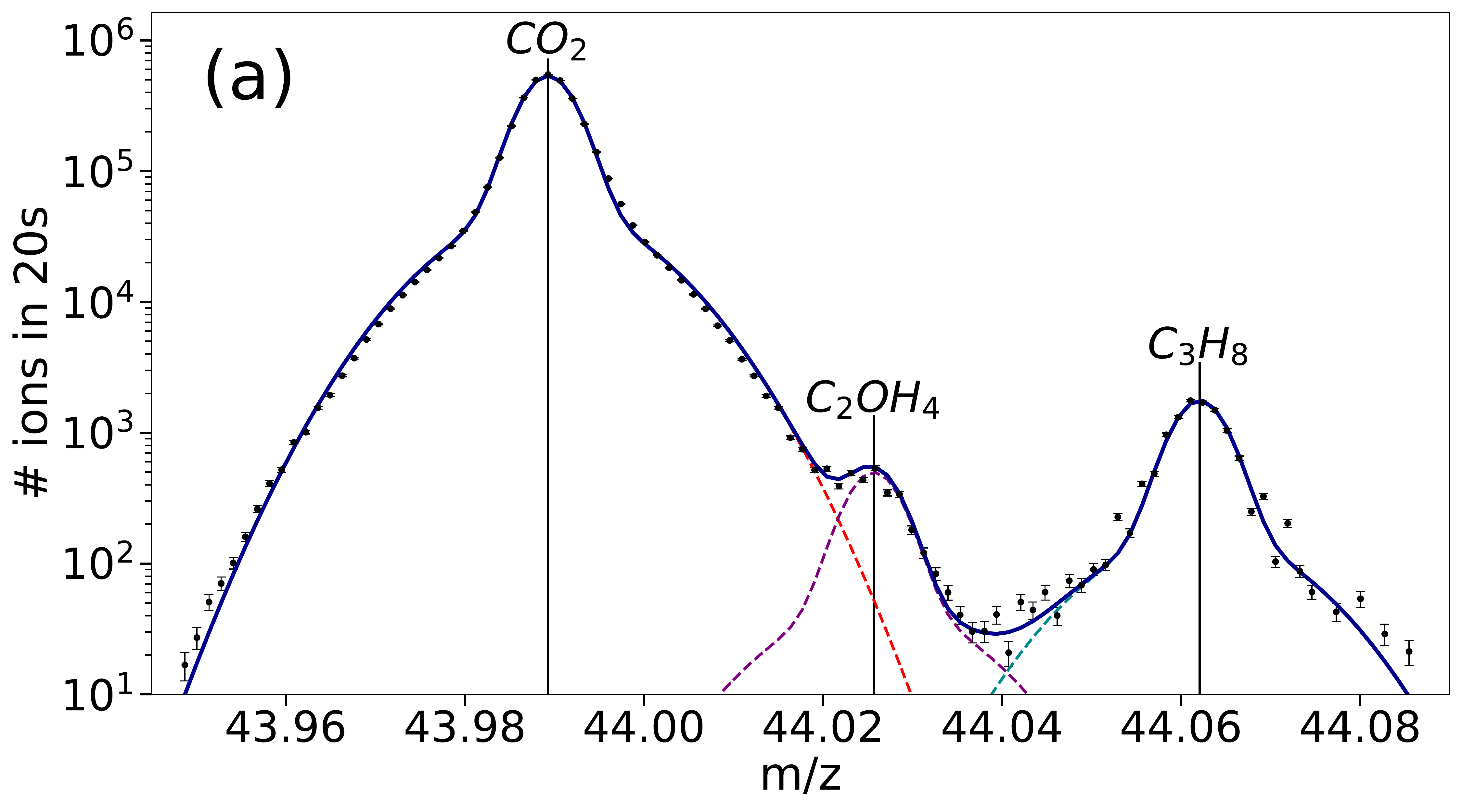}\hfill
    \includegraphics[width=.5\textwidth]{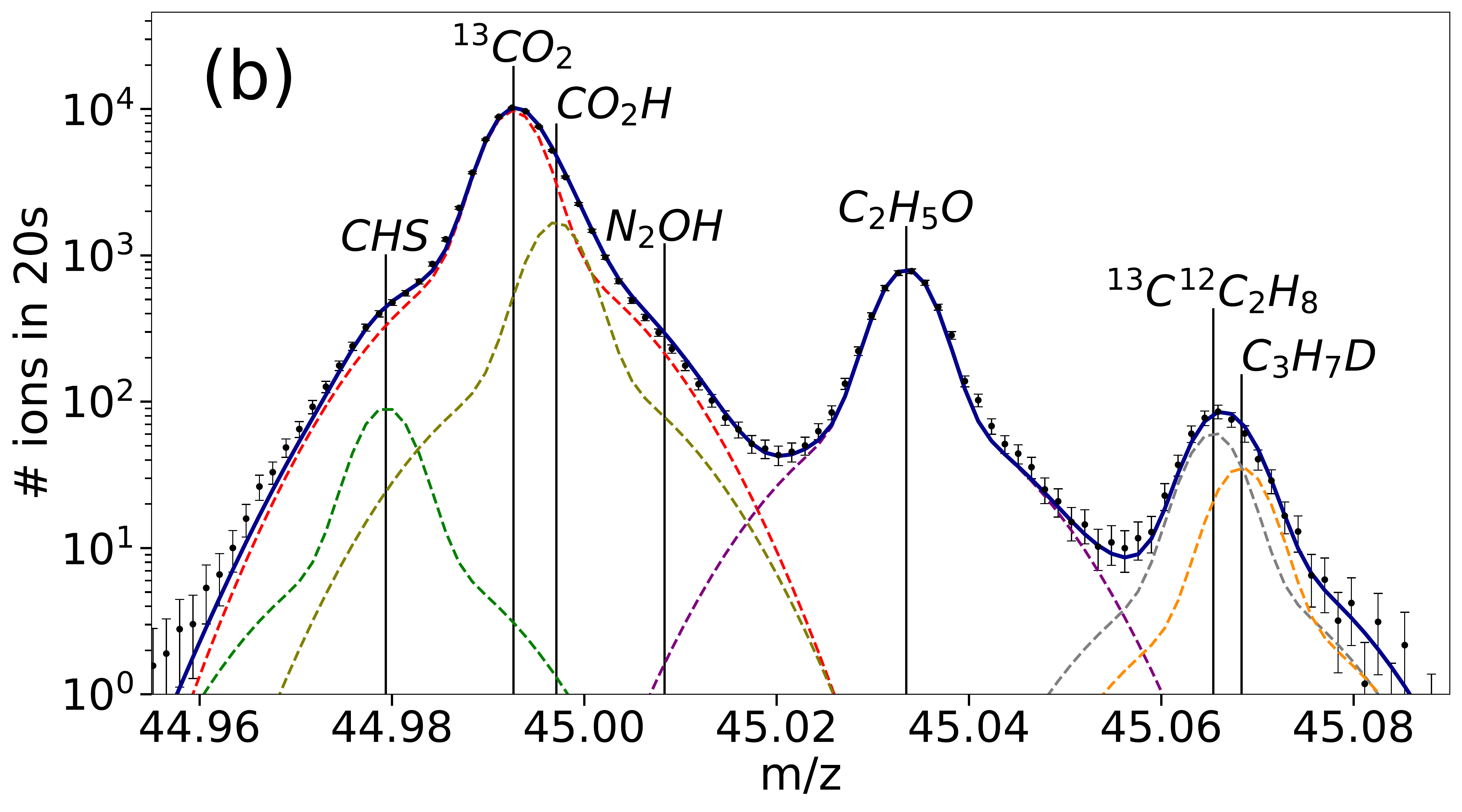}
    \caption{Sample mass spectra for \emph{m/z} 44 and 45 showing the signatures of the isotopologues of propane. \textit{Panel a}: \emph{m/z} 44 from 2016-03-20 15:24 (UTC). \textit{Panel b}: \emph{m/z} 45 from 2016-03-20 15:24 (UTC). Measured data are represented by black dots including their statistical uncertainties. The individual mass fits and the total sum of the fits are shown with coloured lines.}
    \label{fig:sampleSpectraPropane}
\end{figure}

\begin{figure}[h]
    \centering
    \includegraphics[width=.5\textwidth]{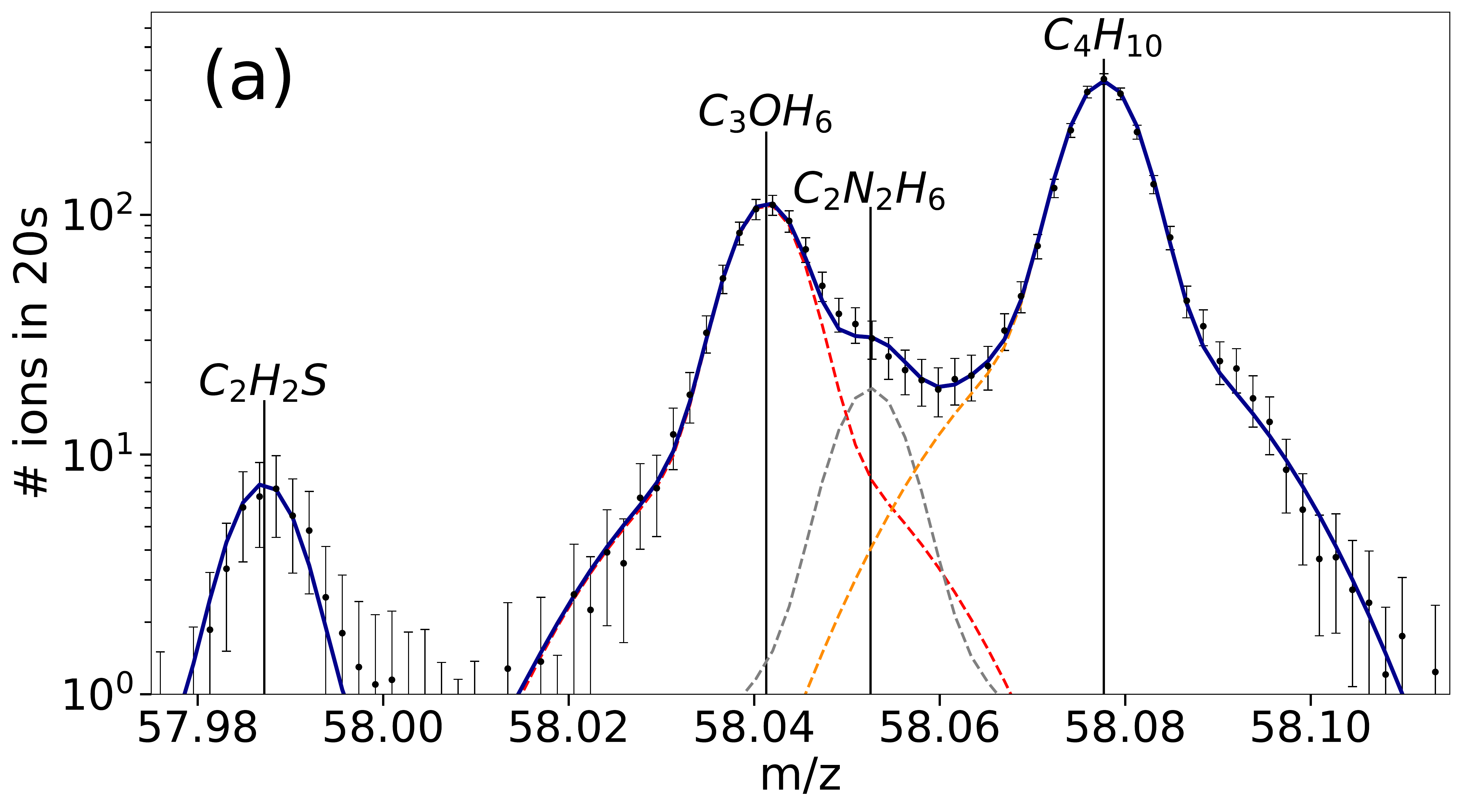}\hfill
    \includegraphics[width=.5\textwidth]{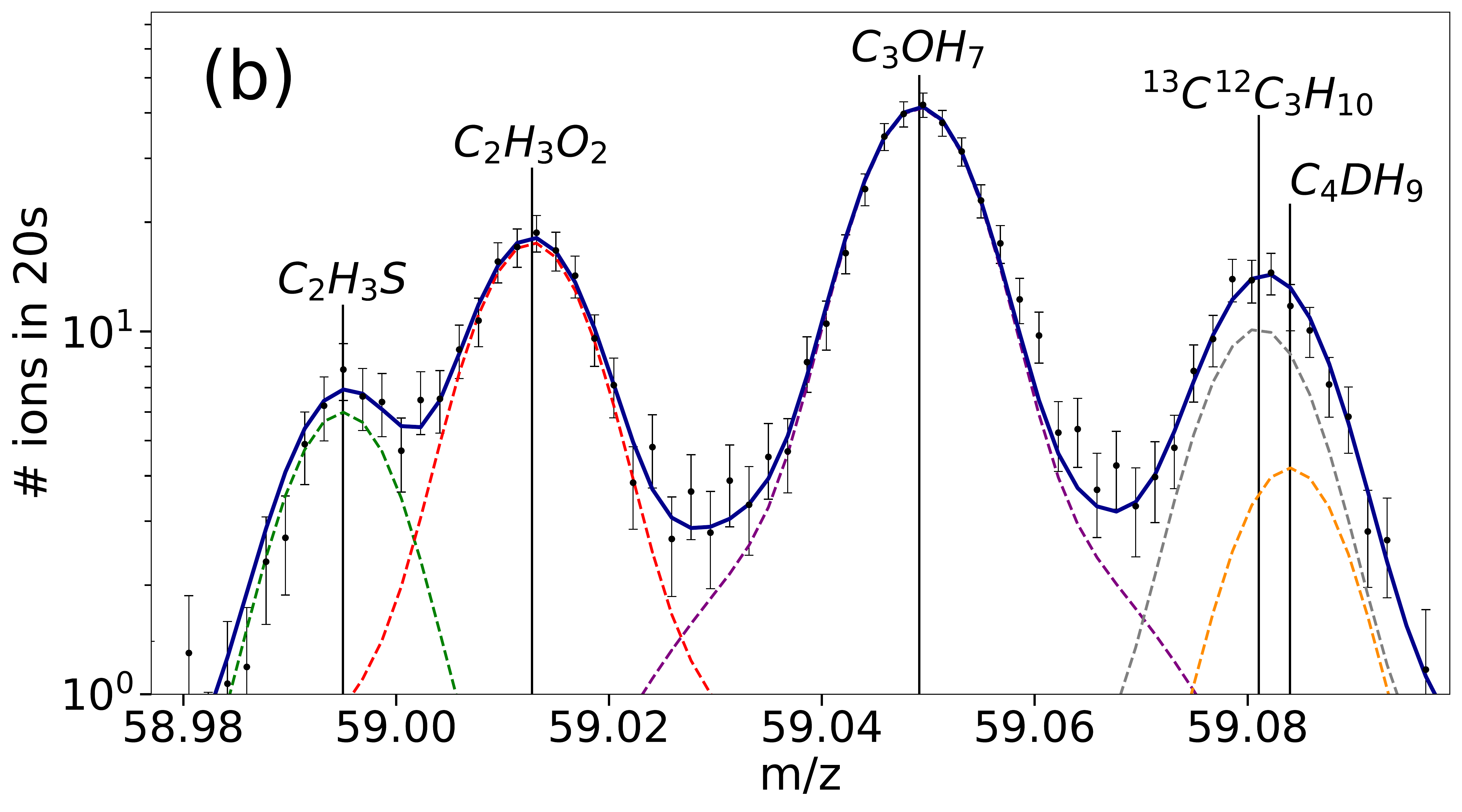}
    \caption{Sample mass spectra for \emph{m/z} 58 and 59 showing the signatures of the isotopologues of butane. \textit{Panel a}: \emph{m/z} 58 from 2016-03-19 22:45 (UTC). \textit{Panel b}: \emph{m/z} 59 from 2016-03-19 22:45 (UTC). Measured data are represented by black dots including their statistical uncertainties. The individual mass fits and the total sum of the fits are shown with coloured lines.}
    \label{fig:sampleSpectraButane}
\end{figure}

\clearpage
\section{Reference tables for literature values used in figures}
\renewcommand{\arraystretch}{0.93} 
\begin{table}[h]
\caption{Literature values for D/H in comets, CCs and IDPs.}
\label{table:LiteratureD}      
\centering          
\begin{tabular}{l l r l}     
\hline
\noalign{\smallskip}
Molecule & Source & D/H & Reference\\
\noalign{\smallskip}
\hline
\noalign{\smallskip}
H\textsubscript{2}O & 67P & (5.01 ± 0.41) · 10\textsuperscript{-4} & This work\\
H\textsubscript{2}O & 103P/Hartley 2 & (1.61 ± 0.24) · 10\textsuperscript{-4} & \citet{Hartogh2011}\\
H\textsubscript{2}O & 46P/Wirtanen & (1.61 ± 0.65) · 10\textsuperscript{-4} & \citet{Lis2019}\\
H\textsubscript{2}O & 81P/Wild 2 & (1.18 - 4.98) · 10\textsuperscript{-4} & \citet{McKeegan2006}\\
H\textsubscript{2}O & C/1995 O1 (Hale-Bopp) & (3.3 ± 0.8) · 10\textsuperscript{-4} & \citet{Meier1998}\\
H\textsubscript{2}O & C/1996 B2 (Hyakutake) & (2.9 ± 1) · 10\textsuperscript{-4} & \citet{Bockelee1998}\\
H\textsubscript{2}O & 8P/Tuttle & (4.09 ± 1.45) · 10\textsuperscript{-4} & \citet{Bockelee2015}\\
H\textsubscript{2}O & C/2009 P1 (Garradd) & (2.06 ± 0.22) · 10\textsuperscript{-4} & \citet{Bockelee2015}\\
H\textsubscript{2}O & C/2014 Q2 (Lovejoy) & (1.4 ± 0.4) · 10\textsuperscript{-4} & \citet{Biver2016}\\
H\textsubscript{2}O & C/2014 Q2 (Lovejoy) & (3.02 ± 0.87) · 10\textsuperscript{-4} & \citet{Paganini2017}\\
H\textsubscript{2}O & CCs/OCs & (0.16 - 5.9) · 10\textsuperscript{-4} & \citet{Alexander2010, Alexander2012}\\
HDO\textsuperscript{a} & 67P & (3.6 ± 1.8) · 10\textsuperscript{-2} & \citet{Altwegg2017}\\
H\textsubscript{2}S & 67P & (1.2 ± 0.3) · 10\textsuperscript{-3} & \citet{Altwegg2017}\\
HCN & C/1995 O1 (Hale-Bopp) & (2.3 ± 0.6) · 10\textsuperscript{-3} & \citet{Crovisier2004}\\
NH\textsubscript{3} & C/1995 O1 (Hale-Bopp) & < 4 · 10\textsuperscript{-2} & \citet{Crovisier2004}\\
NH\textsubscript{3} & 67P & (1.1 ± 0.2) · 10\textsuperscript{-3} & \citet{Altwegg2019}\\
CH\textsubscript{4} & C/2004 Q2 (Machholz) & < 5 · 10\textsuperscript{-3} & \citet{Bonev2009}\\
CH\textsubscript{4} & C/2001 Q4 (NEAT) & < 1 · 10\textsuperscript{-2} & \citet{Kawakita2005}\\
CH\textsubscript{4} & C/2007 N3 (Lulin) & < 7.5 · 10\textsuperscript{-3} & \citet{Gibb2012}\\
CH\textsubscript{4} & 67P & (2.41 ± 0.29) · 10\textsuperscript{-3} & This work\\
C\textsubscript{2}H\textsubscript{6} & C/2007 W1 (Boattini) & < 2.6 · 10\textsuperscript{-3} & \citet{Doney2020}\\
C\textsubscript{2}H\textsubscript{6} & 67P & (2.37 ± 0.27) · 10\textsuperscript{-3} & This work\\
C\textsubscript{3}H\textsubscript{8} & 67P & (2.16 ± 0.26) · 10\textsuperscript{-3} & This work\\
C\textsubscript{4}H\textsubscript{10} & 67P & (2.05 ± 0.38) · 10\textsuperscript{-3} & This work\\
H\textsubscript{2}CO & C/1995 O1 (Hale-Bopp) & < 5 · 10\textsuperscript{-2} & \citet{Crovisier2004}\\
CH\textsubscript{3}OH & C/1995 O1 (Hale-Bopp) & < 3 · 10\textsuperscript{-2} & \citet{Crovisier2004}\\
CH\textsubscript{3}OH & 67P & (0.71 - 6.63) · 10\textsuperscript{-2} & \citet{Drozdovskaya2021}\\
Refr. Organics & 67P & (1.57 ± 0.54) · 10\textsuperscript{-3} & \citet{Paquette2021}\\
Bulk & CCs/OCs & (1.2 - 8.4) · 10\textsuperscript{-3} & \citet{Alexander2010, Alexander2012, Kerridge1985}; \\ & & & \citet{Pearson2001, Yang1984}\\
Organics & CCs/OCs & < 3.1 · 10\textsuperscript{-3} & \citet{Alexander2007, Alexander2010, Busemann2006}\\
Organics & IDPs/UCCAMs & 9 · 10\textsuperscript{-5} - 8 · 10\textsuperscript{-3} & \citet{Duprat2010, Messenger2000}\\
\noalign{\smallskip}
\hline                    
\end{tabular}
\tablebib{
(a)~D/H in HDO is equal to 2 · D\textsubscript{2}O/HDO
}

\vspace*{0.5 cm}
\caption{Literature values for \textsuperscript{13}C/\textsuperscript{12}C in comets, CCs and IDPs.}
\label{table:LiteratureC}      
\centering          
\begin{tabular}{l l r l}     
\hline
\noalign{\smallskip}
Molecule & Source & \textsuperscript{13}C/\textsuperscript{12}C & Reference\\
\noalign{\smallskip}
\hline
\noalign{\smallskip}
C\textsubscript{2} & 4 comets & (1.10 ± 0.12) · 10\textsuperscript{-2} & \citet{Wyckoff2000}\\
CH\textsubscript{4} & 67P & (1.14 ± 0.13) · 10\textsuperscript{-2} & This work\\
C\textsubscript{2}H\textsubscript{6} & 67P & (1.08 ± 0.12) · 10\textsuperscript{-2} & This work\\
C\textsubscript{2}H\textsubscript{6} & 67P & (9.5 ± 0.9) · 10\textsuperscript{-3} & \citet{Altwegg2020}\\
C\textsubscript{3}H\textsubscript{8} & 67P & (1.15 ± 0.12) · 10\textsuperscript{-2} & This work\\
C\textsubscript{4}H\textsubscript{10} & 67P & (1.04 ± 0.15) · 10\textsuperscript{-2} & This work\\
CN & 19 comets & (1.10 ± 0.05) · 10\textsuperscript{-2} & \citet{Manfroid2009}\\
HCN & C/1995 O1 (Hale-Bopp) & (9.0 ± 1.0) · 10\textsuperscript{-3} & \citet{Jewitt1997}\\
CO & 67P & (1.16 ± 0.12) · 10\textsuperscript{-2} & \citet{Rubin2017}\\
CO\textsubscript{2} & 67P & (1.19 ± 0.06) · 10\textsuperscript{-2} & \citet{Haessig2017}\\
H\textsubscript{2}CO & 67P & (2.5 ± 0.9) · 10\textsuperscript{-2} & \citet{Altwegg2020}\\
CH\textsubscript{3}OH & 67P & (1.10 ± 0.12) · 10\textsuperscript{-2} & \citet{Altwegg2020}\\
Bulk & 81P/Wild 2 & (1.1 ± 0.01) · 10\textsuperscript{-2} & \citet{Stadermann2008}\\
Bulk & CCs & (1.09 - 1.15) · 10\textsuperscript{-2} & \citet{Alexander2010, Alexander2012, Pearson2001}\\
Organics & CCs/IDPs & (1.05 - 1.12) · 10\textsuperscript{-2} & \citet{Alexander2007, Floss2004}\\
\noalign{\smallskip}
\hline                    
\end{tabular}
\end{table}

\end{appendix}

\end{document}